\documentclass[11pt,a4paper]{article}

\usepackage{hyperref}
\usepackage[numbers,square,sort&compress]{natbib}
\usepackage{amsfonts}
\usepackage[dvips]{graphicx}
\usepackage{amsmath}
\usepackage{mathrsfs}
\usepackage{makeidx}
\usepackage{xypic}
\usepackage[nottoc]{tocbibind}
\usepackage{pstricks}
\usepackage{bbm}
\usepackage[arrow, matrix, curve]{xy}
\usepackage{amsthm}
\usepackage{amssymb}
\usepackage{epic}
\usepackage{eepic}
\usepackage{times}
\usepackage{epsfig}

\makeatletter
\@addtoreset{equation}{section}
\makeatother

\xymatrixcolsep{0.5cm}
\xymatrixrowsep{0.5cm}

\newtheorem{proposition}{Proposition}[section]
\newtheorem{lemma}{Lemma}[section]

\newtheorem{corollary}{Corollary}[section]

\newcommand{\beqa}{\begin{eqnarray}}
\newcommand{\eeqa}{\end{eqnarray}}

\makeindex

\begin{document}

\begin{flushright}
LPENSL-TH-12-04\\
YITP-SB-12-12
\end{flushright}

\par \vskip .1in \noindent

\vspace{-12pt}

\begin{center}
\begin{LARGE}
\vspace*{1cm}
  {On the form factors of local operators in  the lattice\\ 
  sine-Gordon model}
\end{LARGE}

\vspace{50pt}

\begin{large}

{\bf N.~Grosjean}\footnote[1]{Laboratoire de Physique, UMR 5672
du CNRS, ENS Lyon,  France, nicolas.grosjean@ens-lyon.fr},~~
{\bf J.~M.~Maillet}\footnote[2]{ Laboratoire de Physique, UMR 5672
du CNRS, ENS Lyon,  France,
 maillet@ens-lyon.fr},~~
{\bf G.~Niccoli}\footnote[3]{YITP, Stony Brook University, New York, USA, niccoli@max2.physics.sunysb.edu}
\par

\end{large}

\vspace{50pt}

\today

\vspace{80pt}

\centerline{\bf Abstract} \vspace{1cm}
\parbox{12cm}{\small   We develop a method for computing form factors of local operators in the framework of Sklyanin's separation of variables (SOV) approach to quantum integrable systems. For that purpose, we consider  the sine-Gordon model on a finite lattice and  in finite dimensional cyclic representations as our main example. We first build  our two central tools for computing matrix elements of local operators, namely, a generic  determinant formula for the scalar products of states in the SOV framework and the reconstruction of local fields in terms of the separate variables. The general  form factors are then obtained as sums of  determinants of finite dimensional matrices, their matrix elements being given as  weighted sums running over the separate variables and involving  the Baxter Q-operator eigenvalues. }
\end{center}

\newpage

\tableofcontents

\newpage

\section{Introduction\label{INTR}}

The computation of general matrix elements and correlation functions of
local operators is one of the fundamental problems of quantum field theory
and statistical mechanics. These objects contain indeed the key dynamical
and measurable quantities of the corresponding physical systems, see e.g. 
\cite{VanHo54,VanHo54a,Kub57,KubTH85}. In the integrable (low dimensional)
situation \cite{Heise28,Bet31,Hul38,Orb58,Wal59,LieSM61,LieM66}, thanks to
the existence of powerful algebraic structures related to the Yang-Baxter
algebra (see e.g. \cite%
{Yang67,FadST79,FadT79,Tha81,Bax82L,GauL83,BogIK93L,FadLH96,JimM95L} and
references therein), significant progress towards their exact determination
has been obtained in the last thirty years. Such results concern in
particular models solvable by means of the algebraic Bethe ansatz like the $%
XXZ$ Heisenberg spin chain \cite{Heise28,Bet31,Hul38,Orb58,Wal59}. They were
first obtained at the free fermion point, namely for the Ising model and the
Heisenberg chain (for anisotropy $\Delta$ equal to zero) \cite%
{Mcc68,McCTW77,SatMJ78,McCPW81}. Going beyond such a free fermion case involves a deep
use of the Yang-Baxter algebra. After historical attempts for the finite
chain in the framework of the Bethe ansatz (see \cite{BogIK93L} and
references therein) but leading in fact to implicit representations in terms
of dual fields, explicit representations for form factors and correlation
functions were first obtained directly in the infinite chain (in the massive
regime) \cite{JimMMN92,JimM95L}. The underlying quantum algebra structure
was instrumental there together with a few assumptions on the representation
of the Hamiltonian and of the local spin operators within the representation
theory of this quantum affine algebra. Similar results for the disordered
regime were then derived using the assumption that correlation functions
should satisfy $q$-deformed KZ equations in close analogy with the massive
regime \cite{JimM96}. The derivation of these results (both for massive and
massless regimes and in the presence of a magnetic field) was later obtained
in the framework of the algebraic Bethe ansatz, starting from finite size
systems, thanks to the resolution of the quantum inverse scattering problem 
\cite{KitMT99,KitMT00,MaiT00}. Further investigations led to the extension
of these results to the non-zero temperature case \cite{GohKS04,GohKS05} and
also to the case of non-trivial (integrable) boundary conditions \cite%
{KitKMNST07,KitKMNST08}. The computation of physical (two-point) correlation
functions required sophisticated summation techniques of the elementary
blocks of correlation functions \cite%
{KitMST02a,KitMST05a,KitMST05b,KitKMST07} that ultimately led to the exact
computation of their asymptotic behavior \cite%
{KitKMST09a,KitKMST09b,KozMS11a,KozMS11b}. It is also worth mentioning that
controlled numerical summation techniques combined with these exact results
on form factors led to the determination of dynamical structure factors in
very good agreement with actual neutron scattering experiments on magnetic
crystals, see e.g. \cite{CauM05,CauHM05,PerSCHMWA06,PerSCHMWA07}. Another
important line of research revealed powerful hidden Grassmann structures
that could also be used in the analysis of the conformal limit of these
models \cite%
{BooJMST07,BooJMST09,BooJMS09a,JimMS09,BooJMS09b,JimMS11a,JimMS11b}.
Finally, it was recently shown how to obtain the asymptotic behavior of the
correlation functions for critical systems through their expansion in terms
of form factors in the finite volume \cite{KitKMST11a,KitKMST11b,KitKMST12a}.

For more sophisticated systems, like lattice integrable models related to
higher rank algebras or for integrable relativistic quantum field theories,
the present state of the art is slightly less satisfactory despite
considerable efforts.

On the one hand, the bootstrap approach has provided great insights into the
structure of the exact scattering matrices \cite%
{Zam77,ZamZ78,ZamZ79,KarTTW77,KarT77,BerKKW78,Kar79} and form factors of
such integrable massive quantum field theories \cite%
{KarW78,BerKW79,Smi84,Smi86,Smi92b}. Perturbed conformal field theory%
\footnote{%
See \cite{Vi70,BPZ84,Gin89,Ca88,DFMS97} and references therein for some
literature on conformal field theories.} (CFT) \cite%
{Zam87,Zam88,Zam89,Zam86,Car88,Smi89,ResS90,CarM90,Smi90,AlZam91,Mus92,GM96}
has also been used in this context together with a new understanding of the
integrable structure of CFT \cite{BazLZ96,BazLZ97,BazLZ99}. Several
important attempts have also been made to use Sklyanin's separation of
variable method (SOV) \cite{Skl85,Skl92,Skl95,KuzS98,Skl99,BabBS96,BabBS97}
in these more complicated algebraic situations and in particular in the case
of infinite dimensional representations associated to the quantum fields 
\cite{Skl89,Skl89a,Smi98,Luk01,Bab04,BytT06,Tes08,BytT09}. One should also
mention that the new fermionic structures mentioned above in the case of the 
$XXZ$ lattice model have been used recently to investigate the structure of
matrix elements of the sine-Gordon model in the infinite volume limit \cite%
{JimMS11a,JimMS11b}.

On the other hand, although these advances lead to invaluable informations
on these theories, a direct computation of the form factors and correlation
functions starting from the description of their local fields and using the
given Hamiltonian dynamics is still missing. The main reason is that
although these approaches succeeded to describe more and more efficiently
both the space of states of such models and the space of their form factors,
the connection between the set of operators used to reach this goal and the
local operators of the theory one is interested in has not been found yet.
Consequently, the matrix elements and correlation functions of these local
fields can at best be identified through indirect arguments. In other words,
contrary to the above mentioned finite dimensional cases, the solution of
the quantum inverse scattering problem is in general not known in the field
theory situation; moreover, it appears very often in these more complicated
models (either on the lattice or in the continuum) that the usual algebraic
Bethe ansatz is no longer applicable due to the lack of a proper reference
state, and that the SOV framework \cite{Skl85,Skl92,Skl95} has to be
considered instead. Although this beautiful method is quite general and
powerful to describe the spectrum of these models, the problem of
reconstructing the local operators of the corresponding theories in terms of
the separate variables is in general still open (see however \cite{Bab04}).

To explain in more detail the main issues of this problem, let us consider
in particular the sine-Gordon model which will be the main example analyzed
in this article. As we just recall, integrable massive field theories in
infinite volume can be solved \textit{on-shell}\footnote{%
See for a review \cite{Mus92} and references therein.} through the
determination of their exact S-matrices \cite%
{Zam77,ZamZ78,ZamZ79,KarTTW77,KarT77,BerKKW78,Kar79} which completely
characterize the particle dynamics. In this particle formulation of the
theory a direct access to the local fields is missing and any information
about them needs to be extracted from the particle dynamics. The monodromy
properties and the singularity structure provide a set of functional
equations \cite{KarW78,BerKW79,Smi84,Smi86,Smi92b} for the form factors of
the local fields on asymptotic particle states. These form factor equations
are uniquely fixed by the knowledge of the exact S-matrix and the space of
their solutions is expected to coincide with that of the local operators of
the theory; many results are known on the form factors of local fields, see
for example \cite{Smi89,ResS90,CarM90,Smi90}, \cite%
{FriMS90,FMS93-2,KouM93,AhnDM93,MusS94,Kub94,DelfM95,DelfSC96,BabFKZ99,BabFK02,BabFK02-1}
and references therein for some literature related to sine-Gordon model.
However, it is worth pointing out that the form factor equations only
contain information on symmetry data of the fields (like charges and spin)
and after fixing them to some values, we are still left with an infinite
dimensional space of form factor solutions which should correspond to the
infinite dimensional space of local fields sharing these data. There is a
large literature dedicated to the longstanding problem of the identification
of the local fields in the scattering formulation of quantum field theories.
Different methods have been introduced to address it; one important line of
research is related to the description of massive integrable quantum field
theories as (super-renormalizable) perturbations of conformal field theories
by relevant local fields \cite%
{Zam87,Zam88,Zam89,Zam86,Car88,Smi89,ResS90,CarM90,Smi90,AlZam91,Mus92,GM96}%
. This characterization has led to the expectation that the perturbations do
not change the structure of the local fields in this way leading to the
attempt to classify the local field content of massive theories by that of
the corresponding ultraviolet conformal field theories. The latter issue was
first addressed in \cite{CarM90} in the simplest massive free theory, namely
the Ising model. There a conjecture was introduced defining a correspondence
between mild asymptotic behavior at high energy of form factors and chiral
local fields. Such a conjecture was justified showing the isomorphism of the
space of chiral local fields in the massive and conformal models. The
extension of the chiral isomorphism to interacting massive integrable
theories was done in \cite{Kub95} for several massive deformations of
minimal conformal models, in \cite{Smi96} for the sine-Gordon model and in 
\cite{JMT03} for all its reductions to unitary minimal models\footnote{%
An important role in these studies has been played by the fermionic
representations of the characters, as derived for different classes of
rational conformal field theories in \cite%
{KKMcM93-1,KKMcM93-2,DKKMcM93,FNO92,NRT93,KMcM95,BMc98}.}. The problem to
extend the classification also to non-chiral local fields was analyzed in a
series of works \cite{DN05-1,DN05-2,DN06,DN08}; in particular, for the
massive Lee-Yang model, the first proof that the operator space determined
by the particle dynamics coincides with that prescribed by conformal
symmetry at criticality was given in \cite{DN08}. While these are
indubitably important results on the global structure of the operator space
in massive theories it is worth pointing out that they do not lead to the
full identification of particular local fields\footnote{%
Apart for some of them, like the components of the stress energy tensor,
which can be characterized by physical prescription \cite{DN05-1} and \cite%
{DN06}.}. In \cite{BabBS96} a criterion has been introduced based on the
quasi-classical characterization of the local fields; it has been fully
described in the special cases of the restricted sine-Gordon model at the
reflectionless points for chiral fields and verified on the basis of
counting arguments \cite{BabBS97}.

This makes clear that, in the S-matrix formulation of massive quantum
integrable theories in infinite volume, the main open problem remains the
absence of a direct reconstruction of the local fields. One of our
motivation for the present work is to define an exact setup to solve this
problem for one of the most paradigmatic integrable quantum field theory,
namely the sine-Gordon model. As a first step we will consider the
discretized version of the sine-Gordon field theory on a finite lattice.
Moreover we will simplify its dynamics by considering the case for which the
representation space of the exponents of the field and conjugated momentum
is an arbitrary finite dimensional cyclic representation of the quantum
algebra, namely the case where the parameter q = $e^{i\text{h{\hskip-.2em}%
\llap{\protect\rule[1.1ex]{.325em}{.1ex}}{\hskip.2em}}}$\ is a root of
unity. This case is interesting in its own right \cite{NicT10,Nic10,Nic11}
and we also believe that the treatment of the full continuum theory could
then be reached by taking the needed limits; at least, we expect to be able
to identify the main ingredients and structures necessary to reach this goal
and learn enough to extend it to other models on the lattice or in the
continuum. 

It is worth to describe schematically the microscopic
approach that we intend to follow to solve integrable quantum field
theories by the complete characterization of their spectrum and of their
dynamics. 

The first goal is the solution of the spectral problem, for the lattice and the continuum theories:\\
i) Solution of the spectral problem for the integrable lattice
regularization by the construction of the eigenstates and eigenvalues of the
transfer matrix using the SOV method. \\
ii) Reformulation of the
spectrum in terms of nonlinear integral equations (of thermodynamic Bethe
ansatz type) and definition of finite volume quantum field theories in the 
continuum limit. \\
iii) Derivation of the S matrix and particle description of the
spectrum in the infinite volume (IR) limit. \\
iv) Derivation of the
renormalization group fixed point conformal spectrum in the UV limit. 

The second goal is the solution of the dynamics along the following steps:\\
 i) Determinant formulae for the scalar product
of states in particular involving  transfer matrix eigenstates.\\
ii) Reconstruction of the local operators in terms of
the quantum separate variables.  \\
iii) Computation of matrix elements of local
operators in the eigenstates basis of the transfer matrix. \\
iv) Thermodynamic behavior of the above quantities and computation of 
the physical correlation functions.

In this paper, we develop partly this program for the lattice
regularization of the sine-Gordon model while the completion of it taking
into account the required limits to the continuum theory in finite and
infinite volume case will be addressed in future publications. Let us remark that the
possibility to apply the SOV method for the discretized version of the
sine-Gordon model in finite dimensional cyclic representations was
demonstrated recently in \cite{NicT10}, hence opening the question of the computations of matrix elements of local
fields in a completely controlled (finite dimensional) setting. The first
result of the present paper is to show that the scalar products of states
can be computed in this case as finite dimensional determinants involving in
particular, for eigenstates of the Hamiltonian, the corresponding eigenvalues
of the Baxter Q-operator; orthogonality of different eigenstates can be
proven directly from these expressions. Further, we will show that the
lattice discretization of the local fields of the sine-Gordon model can be
reconstructed explicitly in terms of the separate variables. These two
ingredients finally lead to the determination of the matrix elements of the
exponential of the local fields of the model between arbitrary eigenstates
of the Hamiltonian.

This article is organized as follows. In Section 2 we define the sine-Gordon
model on a finite lattice in the cyclic representations and recall the main
ingredients of the SOV method in that context. In Section 3 we show how to
compute the scalar products of states in the SOV representations. The next
section is devoted to the reconstruction of the local fields in terms of the
separate variables. In Section 5 we use these results to compute the form
factors of the local fields in terms of finite size determinants. In the
last section we comment on these results and compare them to the existing
literature.

\section{The sine-Gordon model}

We use this section to recall the main results derived in \cite{NicT10,Nic10}
on the spectrum description of the lattice sine-Gordon model.

\subsection{Definitions}

\subsubsection{Classical model}

The classical sine-Gordon model can be characterized by the following
Hamiltonian density:%
\begin{equation}
\text{H}_{SG}\equiv \left( \partial _{x}\phi \right) ^{2}+\Pi ^{2}+8\pi \mu
\cos 2\beta \phi
\end{equation}%
where the field $\phi (x,t)$ is defined for $(x,t)\in \lbrack 0,R]\times \ 
\mathbb{R}$ with periodic boundary conditions $\phi (x+R,t)=\phi (x,t)$. The
dynamics of the model in the Hamiltonian formalism is defined in terms of
variables $\phi (x,t),$ $\Pi (x,t)$ with the following Poisson brackets:{}%
\begin{equation}
\{\Pi (x,t),\phi (y,t)\}=2\pi \delta (x-y).
\end{equation}%
The classical integrability of the sine-Gordon model is assured thanks to
the representation of the equation of motion by a zerocurvature condition:%
\begin{equation}
\lbrack \partial _{t}-V(x,t;\lambda ),\partial _{x}-U(x,t;\lambda )]=0,
\end{equation}%
where, by using the Pauli matrices, we have defined:%
\begin{align}
U& =\text{k}_{1}\sigma _{1}\cos \beta \phi +\text{k}_{2}\sigma _{2}\sin
\beta \phi -\text{k}_{3}\sigma _{3}\Pi , \\
V& =-\text{k}_{2}\sigma _{1}\cos \beta \phi -\text{k}_{1}\sigma _{2}\sin
\beta \phi -\text{k}_{3}\sigma _{3}\partial _{x}\phi , \\
\text{k}_{1}& =i\beta \left( \pi \mu \right) ^{1/2}(\lambda -\lambda ^{-1}),%
\text{ \ \ \ k}_{2}=i\beta \left( \pi \mu \right) ^{1/2}(\lambda +\lambda
^{-1}),\text{ k}_{3}\equiv i\frac{\beta }{2}.
\end{align}

\subsubsection{Quantum lattice regularization}

In order to regularize the ultraviolet divergences that arise in the
quantization of the model a lattice discretization is introduced. The field
variables are discretized according to the standard recipe: 
\begin{equation}
\phi _{n}\equiv \phi (n\Delta ),\text{ \ \ \ }\Pi _{n}\equiv \Delta \Pi
(n\Delta ),
\end{equation}%
where $\Delta =R/N$ is the lattice spacing. Then, the canonical quantization
is defined by imposing that $\phi _{n}$\ and $\Pi _{n}$ are self-adjoint
operators satisfying the commutation relations:%
\begin{equation}
\lbrack \phi _{n},\Pi _{m}]=2\pi i\delta _{n,m}.  \label{canonical-CR}
\end{equation}%
The quantum lattice regularization of the sine-Gordon model\footnote{%
The integrability of the time evolution in the lattice sine-Gordon model is
known \cite{IzeK82,BazBR96,FadV92,BobKP93} and the most convenient way to
formulate it uses the Baxter Q-operators; these operators have been
constructed for the closely related chiral Potts model in \cite{Ba08}, see 
\cite{NicT10} for a review and rederivation of these points in the lattice
sine-Gordon model.} that we use here goes back to \cite%
{FadST79,IzeK82,TarTF83} and is related to formulations which have more
recently been studied in \cite{Fad94,FadV94}. Here, the basic operators are
the exponentials of the fields and of the conjugate momenta: 
\begin{equation}
\mathsf{v}_{n}\equiv e^{-i\beta \phi _{n}},\text{ \ \ }\mathsf{u}_{n}\equiv
e^{i\beta \Pi _{n}/2}.
\end{equation}%
Then, the commutation relations (\ref{canonical-CR}) imply that the couples
of unitary operators ($\mathsf{u}_{n}$,$\mathsf{v}_{n}$) generate $\mathsf{N}
$ independent Weyl algebras $\mathcal{W}_{n}$:%
\begin{equation}
\mathsf{u}_{n}\mathsf{v}_{m}=q^{\delta _{nm}}\mathsf{v}_{m}\mathsf{u}_{n}\,,%
\text{ \ \ }q\equiv e^{-i\pi \beta ^{2}},  \label{Weyl}
\end{equation}%
for any $n\in \{1,...,\mathsf{N}\}$. In terms of these basic operators the
quantum lattice sine-Gordon model can be characterized by the following Lax
matrix\footnote{%
Here, we make our analysis for the lattice sine-Gordon model leaving the $%
\kappa _{n}$ and $\xi _{n}$ as free parameters. However, the sine-Gordon
model in the continuum limit is reproduced taking suitable limits on these
parameters by relating them to the mass $\mu $ and the radius $R$ of the
compactified space direction of the model.}: 
\begin{equation}
\mathsf{L}_{n}(\lambda )=\kappa _{n}\left( 
\begin{array}{cc}
\mathsf{u}_{n}(q^{-1/2}\mathsf{v}_{n}\kappa _{n}+q^{1/2}\mathsf{v}%
_{n}^{-1}\kappa _{n}^{-1}) & (\lambda _{n}\mathsf{v}_{n}-(\mathsf{v}%
_{n}\lambda _{n})^{-1})/i \\ 
(\lambda _{n}/\mathsf{v}_{n}-\mathsf{v}_{n}/\lambda _{n})/i & \mathsf{u}%
_{n}^{-1}(q^{1/2}\mathsf{v}_{n}\kappa _{n}^{-1}+q^{-1/2}\mathsf{v}%
_{n}^{-1}\kappa _{n})%
\end{array}%
\right)  \label{Lax}
\end{equation}%
where $\lambda _{n}\equiv \lambda /\xi _{n}$ for any $n\in \{1,...,\mathsf{N}%
\}$ with $\xi _{n}$ and $\kappa _{n}$ parameters of the model.

The monodromy matrix $\mathsf{M}(\lambda )$ is defined by:%
\begin{equation}
\mathsf{M}(\lambda )\equiv \left( 
\begin{array}{cc}
\mathsf{A}(\lambda ) & \mathsf{B}(\lambda ) \\ 
\mathsf{C}(\lambda ) & \mathsf{D}(\lambda )%
\end{array}%
\right) \equiv \mathsf{L}_{\mathsf{N}}(\lambda )\cdots \mathsf{L}%
_{1}(\lambda ),
\end{equation}%
and satisfies the quadratic commutation relations:%
\begin{equation}
R(\lambda /\mu )\,(\mathsf{M}(\lambda )\otimes 1)\,(1\otimes \mathsf{M}(\mu
))\,=\,(1\otimes \mathsf{M}(\mu ))\,(\mathsf{M}(\lambda )\otimes 1)R(\lambda
/\mu )\,,  \label{YBA}
\end{equation}%
with the 6-vertex $R$-matrix: 
\begin{equation}
R(\lambda )=\left( 
\begin{array}{cccc}
q\lambda -q^{-1}\lambda ^{-1} &  &  &  \\[-1mm] 
& \lambda -\lambda ^{-1} & q-q^{-1} &  \\[-1mm] 
& q-q^{-1} & \lambda -\lambda ^{-1} &  \\[-1mm] 
&  &  & q\lambda -q^{-1}\lambda ^{-1}%
\end{array}%
\right) \,.  \label{Rsg}
\end{equation}%
The elements of $\mathsf{M}(\lambda )$ are the generators of the so-called
Yang-Baxter algebra and the commutation relations (\ref{YBA}) imply the
mutual commutativity of the elements of the following one-parameter family
of operators:%
\begin{equation}
\mathsf{T}(\lambda )\equiv \text{tr}_{\mathbb{C}^{2}}\mathsf{M}(\lambda ),
\label{Tdef}
\end{equation}%
the so-called transfer matrix. In the case of a lattice with $\mathsf{N}$
even quantum sites, we can also introduce the operator:%
\begin{equation}
\Theta \equiv \prod_{n=1}^{\mathsf{N}}\mathsf{v}_{n}^{(-1)^{n}},
\label{topological-charge}
\end{equation}%
which plays the role of a \textit{grading operator} in the Yang-Baxter
algebra:

\textbf{Proposition 6 of \cite{NicT10}}\textit{\ For the even chain, the
charge $\Theta $ commutes with the transfer matrix and satisfies the
following commutation relations with the generators of the Yang-Baxter
algebra:%
\begin{eqnarray}
\Theta \mathsf{C}(\lambda ) &=&q\mathsf{C}(\lambda )\Theta \text{, \ \ \ }[%
\mathsf{A}(\lambda ),\Theta ]=0, \\
\mathsf{B}(\lambda )\Theta &=&q\Theta \mathsf{B}(\lambda ),\text{ \ \ }[%
\mathsf{D}(\lambda ),\Theta ]=0.
\end{eqnarray}%
Moreover, the $\Theta $-charge allows to express the asymptotics of the
leading generators $\mathsf{A}(\lambda )$ and $\mathsf{D}(\lambda )$:%
\begin{equation}
\lim_{\log \lambda \rightarrow \mp \infty }\lambda ^{\pm \mathsf{N}}\mathsf{A%
}(\lambda )=\left( \prod_{a=1}^{\mathsf{N}}\frac{\kappa _{a}\xi _{a}^{\pm 1}%
}{i}\right) \Theta ^{\mp 1},\text{ \ \ \ \ \ }\lim_{\log \lambda \rightarrow
\mp \infty }\lambda ^{\pm \mathsf{N}}\mathsf{D}(\lambda )=\left(
\prod_{a=1}^{\mathsf{N}}\frac{\kappa _{a}\xi _{a}^{\pm 1}}{i}\right) \Theta
^{\pm 1}
\end{equation}%
and hence the asymptotic behavior of the transfer matrix:%
\begin{equation}
\lim_{\log \lambda \rightarrow \mp \infty }\lambda ^{\pm \mathsf{N}}\mathsf{T%
}(\lambda )=\left( \prod_{a=1}^{\mathsf{N}}\frac{\kappa _{a}\xi _{a}^{\pm 1}%
}{i}\right) \left( \Theta +\Theta ^{-1}\right) .  \label{asymptotics-T}
\end{equation}
}

\subsection{Cyclic representations\label{Def-cyclic-rep}}

In the present article, we will consider representations where both $\mathsf{%
v}_{m}$ and $\mathsf{u}_{n}\,$ have discrete spectra; in particular, we will
restrict our attention to the case in which $q$ is a root of unity:%
\begin{equation}
\beta ^{2}\,=\,\frac{p^{\prime }}{p}\,,\qquad p,p^{\prime }\in \mathbb{Z}%
^{>0}\,,  \label{beta}
\end{equation}%
with $p$ odd and $p^{\prime }$ even so that $q^{p}=1$. The condition (\ref%
{beta}) implies that the powers $p$ of the generators $\mathsf{u}_{n}$ and $%
\mathsf{v}_{n}$ are central elements of each Weyl algebra $\mathcal{W}_{n}$.
In this case, we can associate a $p$-dimensional linear space R$_{n}$ to any
site $n$ of the chain and we can define on it a cyclic representation of $%
\mathcal{W}_{n}$ as follows:%
\begin{equation}
\mathsf{v}_{n}|k_{n}\rangle =v_{n}q^{k_{n}}|k_{n}\rangle ,\text{ \ }\mathsf{u%
}_{n}|k_{n}\rangle =u_{n}|k_{n}-1\rangle ,\text{\ \ \ \ }\forall k_{n}\in
\{0,...,p-1\},  \label{v-eigenbasis}
\end{equation}%
with the cyclicity condition, 
\begin{equation}
|k_{n}+p\rangle =|k_{n}\rangle .
\end{equation}%
The vectors $|k_{n}\rangle $ define a $\mathsf{v}_{n}$-eigenbasis of the
local space R$_{n}$ and the parameters $v_{n}$ and $u_{n}$ characterize the
central elements of the $\mathcal{W}_{n}$-representation:%
\begin{equation}
\mathsf{v}_{n}^{p}=v_{n}^{p},\text{ \ }\mathsf{u}_{n}^{p}=u_{n}^{p}.
\end{equation}%
Here, we require the unitarity of the generators $\mathsf{u}_{n}$ and $%
\mathsf{v}_{n}$ and the orthonormality of the elements of the $\mathsf{v}%
_{n} $-eigenbasis w.r.t. the scalar product introduced in R$_{n}$.

Let L$_{n}$ be the linear space dual of R$_{n}$ and let $\langle k_{n}|$ be
the elements of the dual basis defined by: 
\begin{equation}
\langle k_{n}|k_{n}^{\prime }\rangle =(|k_{n}\rangle ,|k_{n}^{\prime
}\rangle )\equiv \delta _{k_{n},k_{n}^{\prime }}\text{ \ \ }\forall
k_{n},k_{n}^{\prime }\in \{0,...,p-1\}.
\end{equation}%
From the unitarity of the generators $\mathsf{u}_{n}$ and $\mathsf{v}_{n}$,
the covectors $\langle k_{n}|$ define a $\mathsf{v}_{n}$-eigenbasis in the
dual space L$_{n}$ and the following left representation of Weyl algebra $%
\mathcal{W}_{n}$ :%
\begin{equation}
\langle k_{n}|\mathsf{v}_{n}=v_{n}q^{k_{n}}\langle k_{n}|,\text{ \ }\langle
k_{n}|\mathsf{u}_{n}=u_{n}\langle k_{n}+1|,\text{\ \ \ \ }\forall k_{n}\in
\{0,...,p-1\},
\end{equation}%
with cyclicity condition: 
\begin{equation}
\langle k_{n}|=\langle k_{n}+p|.
\end{equation}

In the \textit{left} and \textit{right} linear spaces:%
\begin{equation}
\mathcal{L}_{\mathsf{N}}\equiv \otimes _{n=1}^{\mathsf{N}}\text{L}_{n},\text{
\ \ \ \ }\mathcal{R}_{\mathsf{N}}\equiv \otimes _{n=1}^{\mathsf{N}}\text{R}%
_{n},
\end{equation}%
the representations of the Weyl algebras $\mathcal{W}_{n}$ induce cyclic
left and right representations of dimension $p^{\mathsf{N}}$ of the
monodromy matrix elements, i.e. cyclic representations of the Yang-Baxter
algebra. Such representations are characterized by the $4\mathsf{N}$
parameters $\kappa =(\kappa _{1},\dots ,\kappa _{\mathsf{N}})$, $\xi =(\xi
_{1},\dots ,\xi _{\mathsf{N}})$, $u=(u_{1},\dots ,u_{\mathsf{N}})$ and $%
v=(v_{1},\dots ,v_{\mathsf{N}})$. The unitarity of the Weyl algebra
generators $\mathsf{u}_{n}$ and $\mathsf{v}_{n}$ implies that the parameters 
$v$ and $u$ are pure phases and the following Hermitian conjugation
properties of the generators of Yang-Baxter algebra hold:

\textbf{Lemma 1 of \cite{Nic10}} \textit{If the parameters of the
representation }$\kappa _{n}^{2}$\textit{\ and }$\xi _{n}^{2}$\textit{\ are
real for any }$n=1,\dots ,\mathsf{N}$\textit{\ and satisfy the constrains}%
\begin{equation}
\varepsilon \equiv -(\kappa _{n}\xi _{n})/\left( \kappa _{n}^{\ast }\xi
_{n}^{\ast }\right) \,\,\,\text{ \textit{is uniform along the chain,}}
\label{cond-T-Normality}
\end{equation}

\textit{then it holds: }%
\begin{equation}
\mathsf{M}(\lambda )^{\dagger }\equiv \left( 
\begin{array}{cc}
\mathsf{A}^{\dagger }(\lambda ) & \mathsf{B}^{\dagger }(\lambda ) \\ 
\mathsf{C}^{\dagger }(\lambda ) & \mathsf{D}^{\dagger }(\lambda )%
\end{array}%
\right) =\left( 
\begin{array}{cc}
\mathsf{D}(\lambda ^{\ast }) & \mathsf{C}(\varepsilon \lambda ^{\ast }) \\ 
\mathsf{B}(\varepsilon \lambda ^{\ast }) & \mathsf{A}(\lambda ^{\ast })%
\end{array}%
\right) ,  \label{Hermit-Monodromy}
\end{equation}

\textit{which, in particular, implies the self-adjointness of the transfer
matrix }$\mathsf{T}(\lambda )$\textit{\ for real }$\lambda $\textit{.}

Let us define the average value $\mathcal{O}$ of the elements of the
monodromy matrix $\mathsf{M}(\lambda )$ as 
\begin{equation}
\mathcal{O}(\Lambda )\,=\,\prod_{k=1}^{p}\mathsf{O}(q^{k}\lambda )\,,\qquad
\Lambda \,=\,\lambda ^{p},  \label{avdef}
\end{equation}%
where $\mathsf{O}$ can be $\mathsf{A}$, $\mathsf{B}$, $\mathsf{C}$ or $%
\mathsf{D}$ and we have to remark that the commutativity of each family of
operators $\mathsf{A}(\lambda )$, $\mathsf{B}(\lambda )$, $\mathsf{C}%
(\lambda )$ and $\mathsf{D}(\lambda )$ implies that the corresponding
average values are functions of $\Lambda $, so that $\mathcal{B}(\Lambda )$, 
$\mathcal{C}(\Lambda )$ are Laurent polynomials of degree $[\mathsf{N}]$
while $\mathcal{A}(\Lambda )$, $\mathcal{D}(\Lambda )$ are Laurent
polynomials of degree $\mathsf{\bar{N}}$ in $\Lambda $, where we are using
the notations: 
\begin{equation}
\text{\ }[\mathsf{N}]\equiv \mathsf{N}-\mathtt{e}_{\mathsf{N}},\text{ \ \ }%
\bar{\mathsf{N}}\equiv \mathsf{N}+\mathtt{e}_{\mathsf{N}}-1\text{, \ \ \ \ }%
\mathtt{e}_{\mathsf{N}}\equiv \left\{ 
\begin{array}{l}
1\text{ for }\mathsf{N}\text{ even,} \\ 
0\text{ for }\mathsf{N}\text{ odd.}%
\end{array}%
\right.
\end{equation}

\begin{proposition}
\label{central} \text{}

\begin{itemize}
\item[a)] The average values $\mathcal{A}(\Lambda )$, $\mathcal{B}(\Lambda )$%
, $\mathcal{C}(\Lambda )$, $\mathcal{D}(\Lambda )$ of the monodromy matrix
elements are central elements. Furthermore, they satisfy the following
relations: 
\begin{equation}
(\mathcal{A}(\Lambda ))^{\ast }\equiv \mathcal{D}(\Lambda ^{\ast }),\ \ \ \
\ (\mathcal{B}(\Lambda ))^{\ast }\equiv \epsilon \mathcal{C}(\Lambda ^{\ast
}),  \label{H-cj-A-D}
\end{equation}%
under complex conjugation in the case of self-adjoint representations.

\item[b)] Let%
\begin{equation*}
\mathcal{M}(\Lambda )\,\equiv \,\left( 
\begin{array}{cc}
\mathcal{A}(\Lambda ) & \mathcal{B}(\Lambda ) \\ 
\mathcal{C}(\Lambda ) & \mathcal{D}(\Lambda )%
\end{array}%
\right)
\end{equation*}%
be the 2$\times $2 matrix whose elements are the average values of the
elements of the monodromy matrix $\mathsf{M}(\lambda )$, then we have, 
\begin{equation*}
\mathcal{M}(\Lambda )\,=\,\mathcal{L}_{\mathsf{N}}(\Lambda )\,\mathcal{L}_{%
\mathsf{N}-1}(\Lambda )\,\dots \,\mathcal{L}_{1}(\Lambda )\,,
\end{equation*}%
where 
\begin{equation}
\mathcal{L}_{n}(\Lambda )\equiv \left( 
\begin{array}{cc}
q^{p/2}u_{n}^{p}(\kappa _{2n}^{p}v_{n}^{p}+v_{n}^{-p}) & \kappa
_{n}^{p}(\Lambda v_{n}^{p}/\xi _{n}^{p}-\xi _{n}^{p}/\Lambda v_{n}^{p})/i^{p}
\\ 
\kappa _{n}^{p}(\Lambda /v_{n}^{p}\xi _{n}^{p}-\xi _{n}^{p}v_{n}^{p}/\Lambda
)/i^{p} & q^{p/2} u_{n}^{-p}(\kappa _{2n}^{p}v_{n}^{-p}+v_{n}^{p})%
\end{array}%
\right) ,  \label{Average-L}
\end{equation}%
is the 2$\times $2 matrix whose elements are the average values of the
elements of the Lax matrix $\mathsf{L}_{n}(\lambda )$.
\end{itemize}
\end{proposition}

A similar statement was first proven in \cite{Ta}. In the present paper, we
will be mainly restricted to the case:%
\begin{equation}
u_{n}=1,\,\,\,v_{n}=1\text{ \ \ for \ }n=1,\dots ,\mathsf{N}.
\label{Special-rep}
\end{equation}%
In these representations it holds:

\begin{lemma}
\label{zeros-B} The power 2p of the zeros of \textsf{$B$}$(\lambda )$ are
real numbers with possible complex conjugate couples.
\end{lemma}

\begin{proof}
The previous proposition and the equality 
\begin{equation}
\mathcal{C}(\Lambda )=\mathcal{B}(\Lambda ),
\end{equation}%
which holds for the sine-Gordon representations with $\mathsf{u}_{n}^{p}=%
\mathsf{v}_{n}^{p}=1$, as proven in \cite{NicT10}, imply that%
\begin{equation}
\left( \mathcal{B}(\Lambda )\right) ^{\ast }=\epsilon\mathcal{B}(\Lambda
^{\ast }),
\end{equation}%
and so the statement of the lemma.
\end{proof}

\subsection{SOV-representations of the Yang-Baxter algebra}

\label{SOV-Left}

According to Sklyanin's method \cite{Skl85,Skl92,Skl95}, a separation of
variables (SOV) representation for the spectral problem of the transfer
matrix $\mathsf{T}(\lambda )$ is defined as a representation where the
commutative family of operators $\mathsf{B}(\lambda )$ is diagonal. In \cite%
{NicT10}, the following theorem has been shown:

\textbf{Theorem 1 of \cite{NicT10}} \textit{For almost all the values of the
parameters }$\kappa $\textit{\ and }$\xi $\textit{\ of the representation,
there exists a SOV representation for the sine-Gordon model, i.e. }$\mathsf{B%
}(\lambda )$\textit{\ is diagonalizable and with simple spectrum.}

The proof of this has been given by a recursive construction of the left
cyclic SOV-representations for the sine-Gordon model. Let us recall here the
left SOV-representations of the generators of the Yang-Baxter algebra. The
Proposition \ref{central} fixes the average values of $\mathsf{B}(\lambda )$:%
\begin{equation}
\mathcal{B}(\Lambda )=\left( \prod_{n=1}^{\mathsf{N}}\frac{\kappa _{n}}{i}%
\right) ^{p}Z_{\mathsf{N}}^{\mathtt{e}_{\mathsf{N}}}\prod_{a=1}^{[\mathsf{N}%
]}(\Lambda /Z_{a}-Z_{a}/\Lambda )
\end{equation}%
in terms of the parameters of the representations. Note that the simplicity
of the spectrum of $\mathsf{B}(\lambda )$ is equivalent to the requirement $%
Z_{a}\neq Z_{b}$ for any $a\neq b\in \{1,\dots ,\mathsf{N}-\mathtt{e}_{%
\mathsf{N}}\}$. Moreover, the reality condition of the polynomial $\mathcal{B%
}(\Lambda )$, proven in Lemma \ref{zeros-B}, implies that we can chose a $%
\mathsf{N}$-tupla $\{\eta _{1}^{(0)},...,\eta _{\mathsf{N}}^{(0)}\}$ of $p$%
-roots of $\{Z_{1},...,Z_{\mathsf{N}}\}$ which satisfy the requirements:%
\begin{equation}
\left( \eta _{a}^{(0)}\right) ^{2}\in \mathcal{R}\text{ or }\left( \eta
_{a}^{(0)}\right) ^{2}\notin \mathcal{R}\rightarrow \text{ }\exists b\in
\{1,...,\mathsf{N}\}\backslash a:\left( \eta _{a}^{(0)}\right) ^{2}=\left(
\left( \eta _{b}^{(0)}\right) ^{2}\right) ^{\ast }\text{.}
\end{equation}%
Let $\langle \eta _{_{1}}^{(k_{1})},...,\eta _{\mathsf{N}}^{(k_{\mathsf{N}%
})}|\in \mathcal{L}_{\mathsf{N}}$ be the generic element of a basis of left 
eigenstates of $\mathsf{B}(\lambda )$:%
\begin{equation}
\langle \mathbf{k}|\mathsf{B}(\lambda )=\mathsf{b}_{\mathbf{k}}(\lambda
)\langle \mathbf{k}|\text{ \ with \ }\mathsf{b}_{\mathbf{k}}(\lambda )\equiv
\left( \prod_{n=1}^{\mathsf{N}}\frac{\kappa _{n}}{i}\right) \eta _{\mathsf{N}%
}^{(k_{\mathsf{N}})\mathtt{e}_{\mathsf{N}}}\prod_{a=1}^{[\mathsf{N}%
]}(\lambda /\eta _{a}^{(k_{a})}-\eta _{a}^{(k_{a})}/\lambda ),
\label{SOV-B-L}
\end{equation}
where we have used the notation%
\begin{equation}
\langle \mathbf{k}|\equiv \langle \eta _{_{1}}^{(k_{1})},...,\eta _{\mathsf{N%
}}^{(k_{\mathsf{N}})}|,\text{ \ \ }\,\eta _{a}^{(k_{a})}\equiv q^{k_{a}}\eta
_{a}^{(0)}\text{ }\forall a\in \{1,\dots ,\mathsf{N}\}\,,\text{ \ \ }\mathbf{%
k}\equiv (k_{1},\dots ,k_{\mathsf{N}})\in \mathbb{Z}_{p}^{\mathsf{N}}\,.
\end{equation}%
Then the left action of the other Yang-Baxter generators reads:%
\begin{align}
\langle \mathbf{k}|\mathsf{A}(\lambda )& =\mathtt{e}_{\mathsf{N}}\frac{%
\mathsf{b}_{\mathbf{k}}(\lambda )}{\eta _{\mathsf{N}}^{(k_{\mathsf{N}})}}(%
\frac{\lambda }{\eta _{\mathbf{k,}\mathsf{A}}}\langle \mathbf{k}|\mathsf{T}_{%
\mathsf{N}}^{-}-\frac{\eta _{\mathbf{k,}\mathsf{A}}}{\lambda }\langle 
\mathbf{k}|\mathsf{T}_{\mathsf{N}}^{+})+\sum_{a=1}^{[\mathsf{N}%
]}\prod_{b\neq a}\frac{(\frac{\lambda }{\eta _{b}^{(k_{b})}}-\frac{\eta
_{b}^{(k_{b})}}{\lambda })}{(\frac{\eta _{a}^{(k_{a})}}{\eta _{b}^{(k_{b})}}-%
\frac{\eta _{b}^{(k_{b})}}{\eta _{a}^{(k_{a})}})}a(\eta
_{a}^{(k_{a})})\langle \mathbf{k}|\mathsf{T}_{a}^{-},  \label{SOV-A-L} \\
&  \notag \\
\langle \mathbf{k}|\mathsf{D}(\lambda )& =\mathtt{e}_{\mathsf{N}}\frac{%
\mathsf{b}_{\mathbf{k}}(\lambda )}{\eta _{\mathsf{N}}^{(k_{\mathsf{N}})}}(%
\frac{\lambda }{\eta _{\mathbf{k,}\mathsf{D}}}\langle \mathbf{k}|\mathsf{T}_{%
\mathsf{N}}^{+}-\frac{\eta _{\mathbf{k,}\mathsf{D}}}{\lambda }\langle 
\mathbf{k}|\mathsf{T}_{\mathsf{N}}^{-})+\sum_{a=1}^{[\mathsf{N}%
]}\prod_{b\neq a}\frac{(\frac{\lambda }{\eta _{b}^{(k_{b})}}-\frac{\eta
_{b}^{(k_{b})}}{\lambda })}{(\frac{\eta _{a}^{(k_{a})}}{\eta _{b}^{(k_{b})}}-%
\frac{\eta _{b}^{(k_{b})}}{\eta _{a}^{(k_{a})}})}d(\eta
_{a}^{(k_{a})})\langle \mathbf{k}|\mathsf{T}_{a}^{+},  \label{SOV-D-L}
\end{align}%
where $\eta _{\mathbf{k,}\mathsf{A}}$ and $\eta _{\mathbf{k,}\mathsf{D}}$
are defined by:%
\begin{equation}
\eta _{\mathbf{k,}\mathsf{A}}\equiv \eta _{\mathbf{k,}\mathsf{D}}\equiv 
\frac{\prod_{n=1}^{\mathsf{N}}\xi _{n}}{\prod_{n=1}^{\mathsf{N-1}}\eta
_{n}^{(k_{n})}},  \label{eta_A}
\end{equation}%
and the shift operators $\mathsf{T}_{n}^{\pm }$ are defined by: 
\begin{equation}
\langle \eta _{_{1}}^{(k_{1})},...,\eta _{n}^{(k_{n})},...,\eta _{\mathsf{N}%
}^{(k_{\mathsf{N}})}|\mathsf{T}_{n}^{\pm }\equiv \langle \eta
_{_{1}}^{(k_{1})},...,\eta _{n}^{(k_{n}\pm 1)},...,\eta _{\mathsf{N}}^{(k_{%
\mathsf{N}})}|.
\end{equation}%
The operator family $\mathsf{C}(\lambda )$ is uniquely\footnote{%
Note that the operator $\mathsf{B}(\lambda )$ is invertible except for $%
\lambda $ which coincides with one of its zeros, so in general $\mathsf{C}%
(\lambda )$ is defined by (\ref{qdetdef}) just inverting $\mathsf{B}(\lambda
)$. This is enough to fix in an unique way $\mathsf{C}(\lambda )$, as it is
a Laurent polynomial of degree [$\mathsf{N}$] in $\lambda $.} defined by the
quantum determinant relation:%
\begin{equation}
\mathrm{det_{q}}\mathsf{M}(\lambda )\,\equiv \,\mathsf{A}(\lambda )\mathsf{D}%
(q^{-1}\lambda )-\mathsf{B}(\lambda )\mathsf{C}(q^{-1}\lambda ),
\label{qdetdef}
\end{equation}%
where $\mathrm{det_{q}}\mathsf{M}(\lambda )$ is a central element\footnote{%
The centrality of the quantum determinant in the Yang-Baxter algebra was
first discovered in \cite{IzeK81}, see also \cite{IzeK09} for an historical
note.} of the Yang-Baxter algebra (\ref{YBA}) which reads:%
\begin{equation}
\mathrm{det_{q}}\mathsf{M}(\lambda )\equiv \prod_{n=1}^{\mathsf{N}}\kappa
_{n}^{2}(\lambda /\mu _{n,+}-\mu _{n,+}/\lambda )(\lambda /\mu _{n,-}-\mu
_{n,-}/\lambda ),  \label{q-det-f}
\end{equation}%
where $\mu _{n,\pm }\equiv i\kappa _{n}^{\pm 1}q^{1/2}\xi _{n}$. In the
representations which satisfy (\ref{Special-rep})\ the coefficients of the
SOV-representations can be fixed by introducing the following Laurent
polynomials:%
\begin{equation}
a(\lambda )=(-i)^{\mathsf{N}}\prod_{n=1}^{\mathsf{N}}\frac{\kappa _{n}}{%
\lambda _{n}}(1+iq^{-1/2}\lambda _{n}\kappa _{n})(1+iq^{-1/2}\lambda
_{n}/\kappa _{n}),\text{ \ \ }d(\lambda )=q^{\mathsf{N}}a(-\lambda q),
\label{a&d-def}
\end{equation}%
Note that these coefficients are related to the quantum determinant by:%
\begin{equation}
\mathrm{det_{q}}\mathsf{M}(\lambda )=a(\lambda )d(\lambda /q).
\label{q-det-ad}
\end{equation}%
Note that for the choice of the parameters $\{\kappa _{n}\}\in i\mathbb{R}^{%
\mathsf{N}}$ and $\{\xi _{n}\}\in \mathbb{R}^{\mathsf{N}}$ the numbers  \textsc{k}$\equiv \prod_{n=1}^{\mathsf{N}}\kappa _{n}/i$ and $\mu _{n,\pm }^{p}$ are real.

\subsection{SOV-characterization of the $\mathsf{T}$-spectrum}

Let us denote by $\Sigma _{\mathsf{T}}$ the set of the eigenvalue functions $%
t(\lambda )$ of the transfer matrix $\mathsf{T}(\lambda )$. From the
definitions (\ref{Lax}) and (\ref{Tdef}), $\Sigma _{\mathsf{T}}$ is a subset
of $\mathbb{R}[\lambda ^{2},\lambda ^{-2}]_{\bar{\mathsf{N}}/2}$ where $%
\mathbb{R}[x,x^{-1}]_{\mathsf{M}}$ denotes the linear space over the field $%
\mathbb{R}$ of the \textit{real} Laurent polynomials of degree $\mathsf{M}$
in the variable $x$: 
\begin{equation}
(f(x))^{\ast }=f(x^{\ast })\,\,\,\forall x\in \mathbb{C} \text{ \ \ with \ }%
f(x)\in \mathbb{R}[x,x^{-1}]_{\mathsf{M}}.
\end{equation}

Note that in the case of $\mathsf{N}$ even, the $\Theta $-charge naturally
induces the grading $\Sigma _{\mathsf{T}}=\bigcup_{k=0}^{(p-1)/2}\Sigma _{%
\mathsf{T}}^{k}$, where: 
\begin{equation}
\Sigma _{\mathsf{T}}^{k}\equiv \left\{ t(\lambda )\in \Sigma _{\mathsf{T}%
}:\lim_{\log \lambda \rightarrow \mp \infty }\lambda ^{\pm \mathsf{N}%
}t(\lambda )=\left( \prod_{a=1}^{\mathsf{N}}\frac{\kappa _{a}\xi _{a}^{\pm 1}%
}{i}\right) (q^{k}+q^{-k})\right\} .  \label{asymptotics-t}
\end{equation}%
This simply follows from the asymptotics of $\mathsf{T}(\lambda )$ and from
its commutativity with $\Theta $.

In the SOV-representations the spectral problem for $\mathsf{T}(\lambda )$
is reduced to the following discrete system of Baxter-like equations in the
wave-function $\Psi _{t}(\mathbf{k})\equiv \langle \mathbf{k}|\,t\,\rangle $
of a $\mathsf{T}$-eigenstate $|\,t\,\rangle $: 
\begin{equation}
t(\eta _{r}^{(k_{r})})\Psi _{t}(\mathbf{k})=a(\eta _{r}^{(k_{r})})\Psi _{t}(%
\mathsf{T}_{r}^{-}(\mathbf{k}))+d(\eta _{r}^{(k_{r})})\Psi _{t}(\mathsf{T}%
_{r}^{+}(\mathbf{k}))\,\qquad \text{ }\forall r\in \{1,...,[\mathsf{N}]\}.
\label{SOVBax1}
\end{equation}%
Here, we have denoted by $\mathsf{T}_{r}^{\pm }(\mathbf{k})\equiv
(k_{1},\dots ,k_{r}\pm 1,\dots ,k_{\mathsf{N}})$ and $a(\eta _{r}^{(k_{r})})$
and $d(\eta _{r}^{(k_{r})})$ the coefficients of the SOV-representation as
defined in (\ref{a&d-def}). In the case of $\mathsf{N}$ even, from the
asymptotics of $\mathsf{T}(\lambda )$ given in (\ref{asymptotics-T}), we
have to add to the system (\ref{SOVBax1}) the following equations in the
variable $\eta _{\mathsf{N}}^{(k_{\mathsf{N}})}$: 
\begin{equation}
\Psi _{t,\pm k}(\mathsf{T}_{\mathsf{N}}^{+}(\mathbf{k}))\,=\,q^{\mp k}\Psi
_{t,\pm k}(\mathbf{k}),  \label{SOVBax2}
\end{equation}%
for the wave function $\Psi _{t,\pm k}(\mathbf{k})\equiv \langle \mathbf{k}%
|\,t_{\pm k}\rangle $, where $|\,t_{\pm k}\rangle $ is a simultaneous
eigenstate of $\mathsf{T}(\lambda )$ and $\Theta $ corresponding to $%
t(\lambda )\in \Sigma _{\mathsf{T}}^{k}$ and $\Theta $-eigenvalue $q^{\pm k}$
with $k\in \{0,...,(p-1)/2\}$.

In the paper \cite{Nic10}, a complete characterization of the $\mathsf{T}$%
-spectrum (eigenvalues and eigenstates) has been given in terms of a certain
class of polynomial solutions of a given functional equation. Let us recall
here these results; to this aim let us introduce the one-parameter family $%
D(\lambda )$ of $p\times p$ matrices: 
\begin{equation}
D(\lambda )\equiv 
\begin{pmatrix}
t(\lambda ) & -d(\lambda ) & 0 & \cdots & 0 & -a(\lambda ) \\ 
-a(q\lambda ) & t(q\lambda ) & -d(q\lambda ) & 0 & \cdots & 0 \\ 
0 & {\quad }\ddots &  &  &  & \vdots \\ 
\vdots &  & \cdots &  &  & \vdots \\ 
\vdots &  &  & \cdots &  & \vdots \\ 
\vdots &  &  &  & \ddots {\qquad } & 0 \\ 
0 & \ldots & 0 & -a(q^{p-2}\lambda ) & t(q^{p-2}\lambda ) & 
-d(q^{p-2}\lambda ) \\ 
-d(q^{p-1}\lambda ) & 0 & \ldots & 0 & -a(q^{p-1}\lambda ) & 
t(q^{p-1}\lambda )%
\end{pmatrix}
\label{D-matrix}
\end{equation}%
where for now $t(\lambda )$ is just a real even Laurent polynomial of degree 
$\mathsf{\bar{N}}$ in $\lambda $. Then the determinant of the matrix $%
D(\lambda )$ is an even Laurent polynomial of maximal degree $\mathsf{\bar{N}%
}$ in $\Lambda \equiv \lambda ^{p}$ and we have the following complete
characterization of the transfer matrix spectrum:

\begin{proposition}
\label{SOV-T-Q}The set $\Sigma _{\mathsf{T}}$ coincides with the set of all
the $t(\lambda )\in \mathbb{R}[\lambda ^{2},\lambda ^{-2}]_{\mathsf{\bar{N}}%
/2}$ solutions of the functional equation: 
\begin{equation}
\det_{p}\text{$D$}(\Lambda )=0,\text{ \ \ }\forall \Lambda \in \mathbb{C}.
\label{I-Functional-eq}
\end{equation}%
Moreover, for $\mathsf{N}$ odd the spectrum of $\mathsf{T}(\lambda )$ is
simple and the eigenstate $|\,t\rangle $ corresponding to $t(\lambda )\in
\Sigma _{\mathsf{T}}$ is characterized by:%
\begin{equation}
\Psi _{t}(\mathbf{k})\equiv \langle \,\eta _{1}^{(k_{1})},...,\eta _{\mathsf{%
N}}^{(k_{\mathsf{N}})}\,|\,t\,\rangle =\prod_{r=1}^{\mathsf{N}}Q_{t}(\eta
_{r}^{(k_{r})}),  \label{Qeigenstate-odd}
\end{equation}%
while for $\mathsf{N}$ even the simultaneous spectrum of $\mathsf{T}(\lambda
)$ and $\Theta $ is simple and the eigenstate $|\,t_{\pm k}\rangle $
corresponding to $t(\lambda )\in \Sigma _{\mathsf{T}}^{k}$ and $\Theta $%
-eigenvalue $q^{\pm k}$ with $k\in \{0,...,(p-1)/2\}$ is characterized by:%
\begin{equation}
\Psi _{t,\pm k}(\mathbf{k})\equiv \langle \eta _{1}^{(k_{1})},...,\eta _{%
\mathsf{N}}^{(k_{\mathsf{N}})}\,|\,t_{\pm k}\rangle =(\eta _{\mathsf{N}%
}^{(k_{\mathsf{N}})})^{\mp k}\prod_{r=1}^{\mathsf{N}-1}Q_{t}(\eta
_{r}^{(k_{r})}).  \label{Qeigenstate-even}
\end{equation}%
Here,%
\begin{align}
Q_{t}(\lambda )& =\lambda ^{a_{t}}\prod_{h=1}^{(p-1)\mathsf{N}%
-(b_{t}+a_{t})}(\lambda _{h}-\lambda ),\,\,\,\,\,\,\,\,0\leq a_{t}\leq
p-1,\,\,0\leq b_{t}\leq (p-1)\mathsf{N},  \label{Q_t-definition} \\
a_{t}& =0,\,\,\,\,\,\,\,\,\,b_{t}=0\,\,\mathsf{mod}\,p,\text{ \ \ \ for }%
\mathsf{N}\text{\textsf{\ }odd}  \label{Q-odd} \\
\text{ }a_{t}& =\pm k\,\,\mathsf{mod}\,p,\,\,\,\,\,\,\,\,\,b_{t}=\pm k\,\,%
\mathsf{mod}\,p,\text{ \ \ \ for }\mathsf{N}\text{\textsf{\ }even
and\thinspace\ }t(\lambda )\in \Sigma _{\mathsf{T}}^{k}\text{, \ }
\label{Q-even}
\end{align}%
is the unique (up to quasi-constants) real polynomial solution of the Baxter
functional equation: 
\begin{equation}
t(\lambda )Q_{t}(\lambda )=a(\lambda )Q_{t}(\lambda q^{-1})+d(\lambda
)Q_{t}(\lambda q)\ \ \ \ \forall \lambda \in \mathbb{C},  \label{EQ-Baxter-R}
\end{equation}%
corresponding to the given $t(\lambda )$, which has been constructed in
terms of the cofactors of the matrix $D$$(\Lambda )$, in Theorems 2 and 3 of 
\cite{Nic10}.
\end{proposition}

\section{Decomposition of the identity in the SOV-basis}

The main goal of this section is to obtain the decomposition of the identity in the SOV basis. This is an important step towards the decomposition of any correlation function in terms of matrix elements of local operators in this basis. In the previous sections we recalled how to define a left SOV basis together with the left action of the Yang-Baxter operators on it. To achieve the identity decomposition we need to define a corresponding right SOV basis. 

Traditionally, the right states are constructed from the left ones by means of hermitian conjugation. In the case of hermitian $\mathsf{B}$ operators, this procedure would lead to right states that are still eigenstates of $\mathsf{B}$ by means of its right action. However, from \eqref{Hermit-Monodromy}, we know that the hermitian conjugate of the $\mathsf{B}$ operator is the operator $\mathsf{C}$ not equal to $\mathsf{B}$. Hence such a procedure (hermitian conjugation of the left SOV basis) would lead to a right SOV basis with respect to the $\mathsf{C}$ operator, namely it would be an eigenstate basis for $\mathsf{C}$.  Although such a route could eventually  lead  to an interesting decomposition of the identity,  it would result in a very complicated structure of the scalar product of states (in particular, left and right states would have no simple orthogonality properties); moreover,  we would not be able to obtain, at least by obvious means, the coefficients in such an identity decomposition  in terms of simple determinants. As a consequence, following this path, the structure of the matrix elements of the local operators would hardly be given in terms of determinants as these are usually obtained as rather simple and localized deformations of the corresponding scalar product formula. 

Therefore, our strategy will be here quite different : we will construct the right SOV basis such that it will be an eigenstate basis for the right action of the $\mathsf{B}$ operators. Hence it will not be obtained as the hermitian conjugate of the left SOV basis; however, due to the simplicity of the $\mathsf{B}$-spectrum, left states will admit natural orthogonality properties with respect to right states. Hence, this procedure will lead to a decomposition of the identity as a single sum over the SOV basis,  with coefficients that are not necessarily positive numbers but that are computable by means of rather simple and universal determinant formula. To show this, we will have to compute the action of left $\mathsf{B}$-eigenstates
(covectors) on right $\mathsf{B}$-eigenstates (vectors); up to an overall constant these actions are completely fixed by
the left and right SOV-representations of the Yang-Baxter algebras when the gauges in the SOV-representations are chosen. 

Let us first define the right
SOV-representation with respect to the $\mathsf{B}$ operator by the following actions:
\begin{align}
\mathsf{B}(\lambda )|\mathbf{k}\rangle & =|\mathbf{k}\rangle \mathsf{b}_{%
\mathbf{k}}(\lambda ),  \label{SOV-B-R} \\
&  \notag \\
\mathsf{A}(\lambda )|\mathbf{k}\rangle & =\mathtt{e}_{\mathsf{N}}(\mathsf{T}%
_{\mathsf{N}}^{+}|\mathbf{k}\rangle \frac{\lambda }{\eta _{\mathbf{k,}%
\mathsf{A}}}-\mathsf{T}_{\mathsf{N}}^{-}|\mathbf{k}\rangle \frac{\eta _{%
\mathbf{k,}\mathsf{A}}}{\lambda })\frac{\mathsf{b}_{\mathbf{k}}(\lambda )}{%
\eta _{\mathsf{N}}^{(k_{\mathsf{N}})}}+\sum_{a=1}^{[\mathsf{N}]}\mathsf{T}%
_{a}^{+}|\mathbf{k}\rangle \prod_{b\neq a}\frac{(\frac{\lambda }{\eta
_{b}^{(k_{b})}}-\frac{\eta _{b}^{(k_{b})}}{\lambda })}{(\frac{\eta
_{a}^{(k_{a})}}{\eta _{b}^{(k_{b})}}-\frac{\eta _{b}^{(k_{b})}}{\eta
_{a}^{(k_{a})}})}\bar{a}(\eta _{a}^{(k_{a})}),  \label{SOV-A-R} \\
&  \notag \\
\mathsf{D}(\lambda )|\mathbf{k}\rangle & =\mathtt{e}_{\mathsf{N}}(\mathsf{T}%
_{\mathsf{N}}^{-}|\mathbf{k}\rangle \frac{\lambda }{\eta _{\mathbf{k,}%
\mathsf{D}}}-\mathsf{T}_{\mathsf{N}}^{+}|\mathbf{k}\rangle \frac{\eta _{%
\mathbf{k,}\mathsf{D}}}{\lambda })\frac{\mathsf{b}_{\mathbf{k}}(\lambda )}{%
\eta _{\mathsf{N}}^{(k_{\mathsf{N}})}}+\sum_{a=1}^{[\mathsf{N}]}\mathsf{T}%
_{a}^{-}|\mathbf{k}\rangle \prod_{b\neq a}\frac{(\frac{\lambda }{\eta
_{b}^{(k_{b})}}-\frac{\eta _{b}^{(k_{b})}}{\lambda })}{(\frac{\eta
_{a}^{(k_{a})}}{\eta _{b}^{(k_{b})}}-\frac{\eta _{b}^{(k_{b})}}{\eta
_{a}^{(k_{a})}})}\bar{d}(\eta _{a}^{(k_{a})}),  \label{SOV-D-R}
\end{align}%
on the generic right $\mathsf{B}$-eigenstate $|\mathbf{k}\rangle \equiv
|\eta _{1}^{(k_{1})},...,\eta _{\mathsf{N}}^{(k_{\mathsf{N}})}\rangle \in 
\mathcal{R}_{\mathsf{N}}$. Here, the coefficients $\bar{a}(\eta _{a})$ and $%
\bar{d}(\eta _{a})$ of the representation are fixed up to the gauge by the
condition:%
\begin{equation}
\mathrm{det_{q}}\mathsf{M}(\lambda )=\bar{d}(\lambda )\bar{a}(\lambda /q);
\end{equation}%
while $\mathsf{C}(\lambda )$ is uniquely defined by the quantum determinant
relation $(\ref{q-det-f})$.

\subsection{Action of left $\mathsf{B}$-eigenstates on right $\mathsf{B}$%
-eigenstates}

It is worth remarking that both for the right and left SOV-representations
we are explicitly asking that the coefficients of the representations of $%
\mathsf{A}(\lambda )$ and $\mathsf{D}(\lambda )$ depend only on the zeros of 
$\mathsf{B}(\lambda )$ on which the corresponding shift operator acts; i.e.
they are separated w.r.t. the zeros of $\mathsf{B}(\lambda )$. Naturally
such requirement is compatible with the Yang-Baxter algebra and the quantum
determinant relation but it is not implied by them. The main point that we
are going to prove here is that this separation requirement completely fixes
the actions of the generic covector in the left SOV-basis on the generic
vector in the right SOV-basis. It may be helpful to explain this last
statement in terms of the properties of the matrices which define the change
of basis to the SOV-basis. Let us define the following isomorphism: 
\begin{equation}
\varkappa :\left( h_{1},...,h_{\mathsf{N}}\right) \in \{1,...,p\}^{\mathsf{N}%
}\rightarrow j=\varkappa \left( h_{1},...,h_{\mathsf{N}}\right) \equiv
h_{1}+\sum_{a=2}^{\mathsf{N}}p^{(a-1)}(h_{a}-1)\in \{1,...,p^{\mathsf{N}}\},
\label{corrisp}
\end{equation}%
then we can write:%
\begin{equation}
\langle \text{\textbf{y}}_{j}|=\langle \text{\textbf{x}}_{j}|U^{(L)}=%
\sum_{i=1}^{p^{\mathsf{N}}}U_{j,i}^{(L)}\langle \text{\textbf{x}}_{i}|\text{
\ \ and\ \ \ }|\text{\textbf{y}}_{j}\rangle =U^{(R)}|\text{\textbf{x}}%
_{j}\rangle =\sum_{i=1}^{p^{\mathsf{N}}}U_{i,j}^{(R)}|\text{\textbf{x}}%
_{i}\rangle ,
\end{equation}%
where we have used the notations:%
\begin{equation}
\langle \text{\textbf{y}}_{j}|\equiv \langle \eta _{1}^{(h_{1})},...,\eta _{%
\mathsf{N}}^{(h_{\mathsf{N}})}|\text{ \ and \ }|\text{\textbf{y}}_{j}\rangle
\equiv |\eta _{1}^{(h_{1})},...,\eta _{\mathsf{N}}^{(h_{\mathsf{N}})}\rangle 
\text{, for }j=\varkappa \left( h_{1},...,h_{\mathsf{N}}\right) \text{,}
\end{equation}%
to represent, respectively, the states of the left and right SOV-basis and:%
\begin{equation}
\langle \text{\textbf{x}}_{j}|\equiv \otimes _{n=1}^{\mathsf{N}}\langle
h_{n}|\text{ \ \ \ \ and \ \ \ }|\text{\textbf{x}}_{j}\rangle \equiv \otimes
_{n=1}^{\mathsf{N}}|h_{n}\rangle \text{, for }j=\varkappa \left(
h_{1},...,h_{\mathsf{N}}\right) \text{,}
\end{equation}%
to represent, respectively, the states of the left and right original $%
\mathsf{v}_{n}$-orthonormal basis. Here, $U^{(L)}$ and $U^{(R)}$ are the $p^{%
\mathsf{N}}\times p^{\mathsf{N}}$ matrices for which it holds:%
\begin{equation}
U^{(L)}\mathsf{B}(\lambda )=\Delta _{\mathsf{B}}(\lambda )U^{(L)},\text{ \ \ 
}\mathsf{B}(\lambda )U^{(R)}=U^{(R)}\Delta _{\mathsf{B}}(\lambda ),
\end{equation}%
where $\Delta _{\mathsf{B}}(\lambda )$ is a diagonal $p^{\mathsf{N}}\times
p^{\mathsf{N}}$ matrix. The diagonalizability and simplicity of the $\mathsf{%
B}$-spectrum imply the invertibility of the matrices $U^{(L)}$ and $U^{(R)}$%
\ and the fact that all the diagonal entries of $\Delta _{\mathsf{B}%
}(\lambda )$ are Laurent polynomials\ in $\lambda $ with different zeros.
Then the following proposition holds:

\begin{proposition}
Right and left SOV-basis are right and left $\mathsf{B}$-eigenbasis such
that the $p^{\mathsf{N}}\times p^{\mathsf{N}}$ matrix:%
\begin{equation}
M\equiv U^{(L)}U^{(R)}
\end{equation}%
is diagonal and characterized by:%
\begin{equation}
M_{jj}=\langle \text{\textbf{y}}_{j}|\text{\textbf{y}}_{j}\rangle =\langle
\eta _{1}^{(h_{1})},...,\eta _{\mathsf{N}}^{(h_{\mathsf{N}})}|\eta
_{1}^{(h_{1})},...,\eta _{\mathsf{N}}^{(h_{\mathsf{N}})}\rangle =\frac{C_{%
\mathsf{N}}\prod_{a=1}^{[\mathsf{N}]}\prod_{k_{a}=1}^{h_{a}}a(\eta
_{a}^{(k_{a})})/\bar{a}(\eta _{a}^{(k_{a}-1)})}{\prod_{1\leq b<a\leq \lbrack 
\mathsf{N}]}(\eta _{a}^{(h_{a})}/\eta _{b}^{(h_{b})}-\eta
_{b}^{(h_{b})}/\eta _{a}^{(h_{a})})},  \label{M_jj}
\end{equation}%
where we have denoted:%
\begin{equation}
j=\varkappa \left( h_{1},...,h_{\mathsf{N}}\right)
\end{equation}%
and $C_{\mathsf{N}}$ is a constant characteristic of the chosen
representations.
\end{proposition}

\begin{proof}
The fact that the matrix $M$ is diagonal is a trivial consequence of the
orthogonality of left and right eigenstates corresponding to different
eigenvalue of $\mathsf{B}(\lambda )$ and of the simplicity of the $\mathsf{B}$-spectrum. 
Indeed, defined $i=\varkappa (k_{1},...,k_{%
\mathsf{N}})$ and $j=\varkappa (h_{1},...,h_{\mathsf{N}})$, the following identities hold:%
\begin{eqnarray}
\mathsf{b}_{\mathbf{k}}(\lambda )M_{ij} &=&\langle \eta
_{1}^{(k_{1})},...,\eta _{\mathsf{N}}^{(k_{\mathsf{N}})}|\mathsf{B}(\lambda
)|\eta _{1}^{(h_{1})},...,\eta _{\mathsf{N}}^{(h_{\mathsf{N}})}\rangle ,
\label{Ortho-0} \\
\mathsf{b}_{\mathbf{h}}(\lambda )M_{ij} &=&\langle \eta
_{1}^{(k_{1})},...,\eta _{\mathsf{N}}^{(k_{\mathsf{N}})}|\mathsf{B}(\lambda
)|\eta _{1}^{(h_{1})},...,\eta _{\mathsf{N}}^{(h_{\mathsf{N}})}\rangle ,
\label{Ortho-1}
\end{eqnarray}%
where the identity $\left( \ref{Ortho-0}\right) $ follows acting with $%
\mathsf{B}(\lambda )$ on the left while the identity $\left( \ref{Ortho-1}%
\right) $ follows acting with $\mathsf{B}(\lambda )$ on the right. Then in
the case $(k_{1},...,k_{\mathsf{N}})\neq (h_{1},...,h_{\mathsf{N}})$ the
identities $\left( \ref{Ortho-0}\right) $ and $\left( \ref{Ortho-1}\right) $
imply that $M_{ij}=0$ for any $i\neq j\in \{1,...,p^{\mathsf{N}}\}.$

Now, let us compute the matrix element $\theta _{a}\equiv \langle \eta
_{_{1}}^{(0)},...,\eta _{a}^{(1)},...,\eta _{\mathsf{N}}^{(0)}|\mathsf{A}%
(\eta _{a}^{(1)})|\eta _{_{1}}^{(0)},...,\eta _{a}^{(0)},...,\eta _{\mathsf{N%
}}^{(0)}\rangle $, where $a\in \{1,...,[\mathsf{N}]\}$. Then using the left
action of the operator $\mathsf{A}(\bar{\eta}_{a})$ we get:%
\begin{equation}
\theta _{a}=a(\eta _{a}^{(1)})\langle \eta _{_{1}}^{(0)},...,\eta
_{a}^{(0)},...,\eta _{\mathsf{N}}^{(0)}|\eta _{_{1}}^{(0)},...,\eta
_{a}^{(0)},...,\eta _{\mathsf{N}}^{(0)}\rangle
\end{equation}%
while using the right action of the operator $\mathsf{A}(\bar{\eta}_{a})$
and the orthogonality of right and left $\mathsf{B}$-eigenstates
corresponding to different eigenvalues we get:%
\begin{equation}
\theta _{a}=\prod_{b\neq a,b=1}^{[\mathsf{N}]}\frac{(\eta _{a}^{(1)}/\eta
_{b}^{(0)}-\eta _{b}^{(0)}/\eta _{a}^{(1)})}{(\eta _{a}^{(0)}/\eta
_{b}^{(0)}-\eta _{b}^{(0)}/\eta _{a}^{(0)})}\bar{a}(\eta _{a}^{(0)})\langle
\eta _{_{1}}^{(0)},...,\eta _{a}^{(1)},...,\eta _{\mathsf{N}}^{(0)}|\eta
_{_{1}}^{(0)},...,\eta _{a}^{(1)},...,\eta _{\mathsf{N}}^{(0)}\rangle
\end{equation}%
and so:%
\begin{equation}
\frac{\langle \eta _{_{1}}^{(0)},...,\eta _{a}^{(1)},...,\eta _{\mathsf{N}%
}^{(0)}|\eta _{_{1}}^{(0)},...,\eta _{a}^{(1)},...,\eta _{\mathsf{N}%
}^{(0)}\rangle }{\langle \eta _{_{1}}^{(0)},...,\eta _{a}^{(0)},...,\eta _{%
\mathsf{N}}^{(0)}|\eta _{_{1}}^{(0)},...,\eta _{a}^{(0)},...,\eta _{\mathsf{N%
}}^{(0)}\rangle }=\frac{a(\eta _{a}^{(1)})}{\bar{a}(\eta _{a}^{(0)})}%
\prod_{b\neq a,b=1}^{[\mathsf{N}]}\frac{(\eta _{a}^{(0)}/\eta
_{b}^{(0)}-\eta _{b}^{(0)}/\eta _{a}^{(0)})}{(\eta _{a}^{(1)}/\eta
_{b}^{(0)}-\eta _{b}^{(1)}/\eta _{a}^{(0)})}.  \label{F1}
\end{equation}%
Then by applying (\ref{F1}) $h_{a}$ times, we get:%
\begin{equation}
\frac{\langle \eta _{_{1}}^{(0)},...,\eta _{a}^{(h_{a})},...,\eta _{\mathsf{N%
}}^{(0)}|\eta _{_{1}}^{(0)},...,\eta _{a}^{(h_{a})},...,\eta _{\mathsf{N}%
}^{(0)}\rangle }{\langle \eta _{_{1}}^{(0)},...,\eta _{a}^{(0)},...,\eta _{%
\mathsf{N}}^{(0)}|\eta _{_{1}}^{(0)},...,\eta _{a}^{(0)},...,\eta _{\mathsf{N%
}}^{(0)}\rangle }=\prod_{k_{a}=1}^{h_{a}}\frac{a(\eta _{a}^{(k_{a})})}{\bar{a%
}(\eta _{a}^{(k_{a}-1)})}\prod_{b\neq a,b=1}^{[\mathsf{N}]}\frac{(\eta
_{a}^{(0)}/\eta _{b}^{(0)}-\eta _{b}^{(0)}/\eta _{a}^{(0)})}{(\eta
_{a}^{(h_{a})}/\eta _{b}^{(0)}-\eta _{b}^{(h_{a})}/\eta _{a}^{(0)})}.
\label{F2}
\end{equation}%
Now, let us consider explicitly the case of even $\mathsf{N}$. In this case
we still have to define the recurrence for $a=\mathsf{N}$. We compute the
matrix element:%
\begin{equation}
\theta _{\mathsf{N}}(\lambda )\equiv \langle \eta _{_{1}}^{(0)},...,\eta _{%
\mathsf{N}-1}^{(0)},\eta _{\mathsf{N}}^{(1)}|\mathsf{A}(\lambda )|\eta
_{1}^{(0)},...,\eta _{\mathsf{N}-1}^{(0)},\eta _{\mathsf{N}}^{(0)}\rangle ,
\end{equation}%
acting with $\mathsf{A}(\lambda )$\ on the right we get:%
\begin{equation}
\theta _{\mathsf{N}}(\lambda )\equiv \frac{\lambda \mathsf{b}_{\mathbf{k}%
}(\lambda )}{\eta _{\mathsf{N}}^{(k_{\mathsf{N}})}\eta _{\mathbf{k,}\mathsf{A%
}}}\langle \eta _{_{1}}^{(0)},...,\eta _{\mathsf{N}-1}^{(0)},\eta _{\mathsf{N%
}}^{(1)}|\eta _{_{1}}^{(0)},...,\eta _{\mathsf{N}-1}^{(0)},\eta _{\mathsf{N}%
}^{(1)}\rangle ,
\end{equation}%
while acting on the left we get:%
\begin{equation}
\theta _{\mathsf{N}}(\lambda )\equiv \frac{\lambda \mathsf{b}_{\mathbf{k}%
}(\lambda )}{\eta _{\mathsf{N}}^{(k_{\mathsf{N}})}\eta _{\mathbf{k,}\mathsf{A%
}}}\langle \eta _{_{1}}^{(0)},...,\eta _{\mathsf{N}-1}^{(0)},\eta _{\mathsf{N%
}}^{(1)}|\eta _{_{1}}^{(0)},...,\eta _{\mathsf{N}-1}^{(0)},\eta _{\mathsf{N}%
}^{(1)}\rangle ,
\end{equation}%
which simply implies:%
\begin{equation}
\langle \eta _{_{1}}^{(0)},...,\eta _{\mathsf{N}-1}^{(0)},\eta _{\mathsf{N}%
}^{(1)}|\eta _{_{1}}^{(0)},...,\eta _{\mathsf{N}-1}^{(0)},\eta _{\mathsf{N}%
}^{(1)}\rangle =\langle \eta _{_{1}}^{(0)},...,\eta _{\mathsf{N}%
-1}^{(0)},\eta _{\mathsf{N}}^{(0)}|\eta _{_{1}}^{(0)},...,\eta _{\mathsf{N}%
-1}^{(0)},\eta _{\mathsf{N}}^{(0)}\rangle .
\end{equation}%
The previous formula implies, for both $\mathsf{N}$ even and odd:%
\begin{equation}
\frac{\langle \eta _{_{1}}^{(h_{a})},...,\eta _{a}^{(h_{a})},...,\eta _{%
\mathsf{N}}^{(h_{\mathsf{N}})}|\eta _{_{1}}^{(h_{a})},...,\eta
_{a}^{(h_{a})},...,\eta _{\mathsf{N}}^{(h_{\mathsf{N}})}\rangle }{\langle
\eta _{_{1}}^{(0)},...,\eta _{a}^{(0)},...,\eta _{\mathsf{N}}^{(0)}|\eta
_{_{1}}^{(0)},...,\eta _{a}^{(0)},...,\eta _{\mathsf{N}}^{(0)}\rangle }%
=\prod_{a=1}^{[\mathsf{N}]}\prod_{k_{a}=1}^{h_{a}}\frac{a(\eta
_{a}^{(k_{a})})}{\bar{a}(\eta _{a}^{(k_{a}-1)})}\prod_{1\leq b<a\leq \lbrack 
\mathsf{N}]}\frac{(\frac{\eta _{a}^{(0)}}{\eta _{b}^{(0)}}-\frac{\eta
_{b}^{(0)}}{\eta _{a}^{(0)}})}{(\frac{\eta _{a}^{(k_{a})}}{\eta
_{b}^{(k_{b})}}-\frac{\eta _{b}^{(k_{b})}}{\eta _{a}^{(k_{a})}})}.
\label{F3}
\end{equation}%
To prove it let us observe that (\ref{F2}) coincides with (\ref{F3}) for $%
h_{2}=...=h_{\mathsf{N}}=0$. Now in the case $h_{3}=...=h_{\mathsf{N}}=0$
with $h_{1}\neq 0$ and $h_{2}\neq 0$, we get (\ref{F3}) by the following
factorization:%
\begin{align}
\frac{\langle \eta _{_{1}}^{(h_{1})},\eta _{2}^{(h_{2})},...,\eta _{\mathsf{N%
}}^{(h_{\mathsf{N}})}|\eta _{_{1}}^{(h_{1})},\eta _{2}^{(h_{2})},...,\eta _{%
\mathsf{N}}^{(h_{\mathsf{N}})}\rangle }{\langle \eta _{_{1}}^{(0)},\eta
_{2}^{(0)},...,\eta _{\mathsf{N}}^{(0)}|\eta _{_{1}}^{(0)},\eta
_{2}^{(0)},...,\eta _{\mathsf{N}}^{(0)}\rangle }& =\frac{\langle \eta
_{_{1}}^{(h_{1})},\eta _{2}^{(0)},...,\eta _{\mathsf{N}}^{(h_{\mathsf{N}%
})}|\eta _{_{1}}^{(h_{1})},\eta _{2}^{(0)},...,\eta _{\mathsf{N}}^{(h_{%
\mathsf{N}})}\rangle }{\langle \eta _{_{1}}^{(0)},\eta _{2}^{(0)},...,\eta _{%
\mathsf{N}}^{(0)}|\eta _{_{1}}^{(0)},\eta _{2}^{(0)},...,\eta _{\mathsf{N}%
}^{(0)}\rangle }  \notag \\
& \times \frac{\langle \eta _{_{1}}^{(h_{1})},\eta _{2}^{(h_{2})},...,\eta _{%
\mathsf{N}}^{(h_{\mathsf{N}})}|\eta _{_{1}}^{(h_{1})},\eta
_{2}^{(h_{2})},...,\eta _{\mathsf{N}}^{(h_{\mathsf{N}})}\rangle }{\langle
\eta _{_{1}}^{(h_{1})},\eta _{2}^{(0)},...,\eta _{\mathsf{N}}^{(0)}|\eta
_{_{1}}^{(h_{1})},\eta _{2}^{(0)},...,\eta _{\mathsf{N}}^{(0)}\rangle } 
\notag \\
& =\prod_{a=1}^{2}\prod_{k_{a}=1}^{h_{a}}\frac{a(\eta _{a}^{(k_{a})})}{\bar{a%
}(\eta _{a}^{(k_{a}-1)})}\prod_{b=2}^{[\mathsf{N}]}\frac{(\eta
_{1}^{(0)}/\eta _{b}^{(0)}-\eta _{b}^{(0)}/\eta _{1}^{(0)})}{(\eta
_{1}^{(h_{1})}/\eta _{b}^{(0)}-\eta _{b}^{(0)}/\eta _{1}^{(h_{1})})}  \notag
\\
& \times \frac{(\eta _{2}^{(0)}/\eta _{1}^{(h_{1})}-\eta _{1}^{(h_{1})}/\eta
_{2}^{(0)})}{(\eta _{2}^{(h_{2})}/\eta _{1}^{(h_{1})}-\eta
_{1}^{(h_{1})}/\eta _{2}^{(h_{2})})}\prod_{b=3}^{[\mathsf{N}]}\frac{(\eta
_{2}^{(0)}/\eta _{b}^{(0)}-\eta _{b}^{(0)}/\eta _{2}^{(0)})}{(\eta
_{2}^{(h_{2})}/\eta _{b}^{(0)}-\eta _{b}^{(0)}/\eta _{2}^{(h_{2})})},
\end{align}%
and so on for the generic case. Finally, from (\ref{F3}) the statement of
the proposition follows being by definition:%
\begin{equation}
\frac{M_{ji}}{\langle \text{\textbf{y}}_{p^{\mathsf{N}}}|\text{\textbf{y}}%
_{p^{\mathsf{N}}}\rangle }\equiv \delta _{i,j}\frac{\langle \eta
_{_{1}}^{(h_{1})},\eta _{2}^{(h_{2})},...,\eta _{\mathsf{N}}^{(h_{\mathsf{N}%
})}|\eta _{_{1}}^{(h_{1})},\eta _{2}^{(h_{2})},...,\eta _{\mathsf{N}}^{(h_{%
\mathsf{N}})}\rangle }{\langle \eta _{_{1}}^{(0)},\eta _{2}^{(0)},...,\eta _{%
\mathsf{N}}^{(0)}|\eta _{_{1}}^{(0)},\eta _{2}^{(0)},...,\eta _{\mathsf{N}%
}^{(0)}\rangle }.
\end{equation}
\end{proof}

\textbf{Remark 2.} The diagonal matrix $M$ is mainly fixed by the
requirement that the left and right representations have a separate form in
the zeros of $\mathsf{B}(\lambda )$. Indeed, this fixes completely the
denominator in all the entries of $M$. Moreover, the constant $C_{\mathsf{N}%
} $ is a function of the representation only via the central elements $%
(Z_{1},...,Z_{\mathsf{N}})$. In the following, we will fix:%
\begin{equation}
C_{\mathsf{N}}\equiv (\eta _{\mathsf{N}}^{(0)}p^{1/2})^{\mathtt{e}_{\mathsf{N%
}}},
\end{equation}%
this choice just amounts to an overall renormalization of the states.
Finally, the products of $a(\lambda )/\bar{a}(\lambda q^{-1})$ in the
numerator of each matrix element is fixed from the choice of the gauge done
in the SOV-representations. Note that we are always free to take the
following gauge:%
\begin{equation}
\bar{a}(\lambda q^{-1})\equiv a(\lambda ),  \label{L-R-gauge}
\end{equation}%
for which the numerator in (\ref{M_jj}) is $C_{\mathsf{N}}$.

\subsection{SOV-decomposition of the identity}

The previous results allow to write the following spectral decomposition of
the identity $\mathbb{I}$:%
\begin{equation}
\mathbb{I}\equiv \sum_{i=1}^{p^{\mathsf{N}}}\mu _{i}|\text{\textbf{y}}%
_{i}\rangle \langle \text{\textbf{y}}_{i}|,
\end{equation}%
in terms of the left and right SOV-basis. Here,%
\begin{equation}
\mu _{i}\equiv \frac{1}{\langle \text{\textbf{y}}_{i}|\text{\textbf{y}}%
_{i}\rangle },
\end{equation}%
is the required analogue of Sklyanin's \textit{measure}\footnote{%
Note that here it cannot be called a measure as the $\mathsf{B}%
(\lambda )$ operator is not self-adjoint, meaning that in general the $\mu_i$ are not positive real  numbers. However the above procedure indeed defines a proper decomposition of the identity operator with simple and computable coefficients.}, which is discrete for the
cyclic representations of the sine-Gordon model. Explicitly, the
SOV-decomposition of the identity reads: 
\begin{equation}
\mathbb{I}\equiv \sum_{h_{1},...,h_{\mathsf{N}}=1}^{p}\prod_{1\leq b<a\leq
\lbrack \mathsf{N}]}((\eta _{a}^{(h_{a})})^{2}-(\eta _{b}^{(h_{b})})^{2})%
\frac{|\eta _{_{1}}^{(h_{1})},...,\eta _{\mathsf{N}}^{(h_{\mathsf{N}%
})}\rangle \langle \eta _{_{1}}^{(h_{1})},...,\eta _{\mathsf{N}}^{(h_{%
\mathsf{N}})}|}{C_{\mathsf{N}}\prod_{b=1}^{[\mathsf{N}]}\omega _{b}(\eta
_{b}^{(h_{b})})},
\end{equation}%
where the separate functions $\omega _{b}(\eta _{b}^{(h_{b})})$ reproduce
the numerator in (\ref{M_jj}):%
\begin{equation}
\omega _{b}(\eta _{b}^{(h_{b})})\equiv \left( \eta _{b}^{(h_{b})}\right) ^{[%
\mathsf{N}]-1}\prod_{k_{b}=1}^{h_{b}}a(\eta _{b}^{(k_{b})})/\bar{a}(\eta
_{b}^{(k_{b}-1)})
\end{equation}%
and they are gauge dependent parameters.

\textbf{Remark 3.} \ Sklyanin's measure\footnote{%
See also \cite{Smi98} for further discussions on the measure.} has been
first introduced by Sklyanin in his article on quantum Toda chain \cite%
{Skl85}. There, it has been derived as a consequence of the self-adjointness
of the transfer matrix w.r.t. the scalar product. In particular, the
Hermitian properties of the operator zeros and their conjugate shift
operators have been fixed to assure the self-adjointness of the transfer
matrix. In the similar but more involved sinh-Gordon
model \cite{BytT06}, the problem related to the uniqueness of the definition
of this measure has been analyzed. There, it has been proven that the
measure is in fact uniquely determined once the positive self-adjointness of
the generators $\mathsf{A}(\lambda )$ and $\mathsf{D}(\lambda )$ is
required. In the compact case of the sine-Gordon model the method used here
insures that the analogue of Sklyanin's measure is uniquely determined up to an overall
constant and the choice of gauge, as discussed in the previous remark. Let
us mention that the Sklyanin's measure has already been derived in \cite%
{GIPST07} for cyclic representations of the related $\tau _{2}$-model%
\footnote{%
The SOV analysis for these representations has been first developed in \cite%
{GIPS06}.} \cite{BS,BBP,B04}. There the measure has been obtained by a
different approach, i.e. by a recursive construction which needs to go
through the recursion in the construction of left and right SOV-basis. It is
interesting to remark that in our purely algebraic derivation we skip these
model dependent features so that the approach we used is suitable for
general compact SOV-representations of the 6-vertex Yang-Baxter algebra.

\subsection{SOV-representation of left and right $\mathsf{T}$-eigenstates 
\label{SOV-T-eigenstates}}

Up to an overall normalization, the SOV-decomposition of the identity and
the SOV-characterization of the transfer matrix spectrum lead to the
following representations of the right eigenstate of the transfer matrix $%
\mathsf{T}(\lambda )$:%
\begin{align}
|t_{\pm k}\rangle & =\sum_{i=1}^{p^{\mathsf{N}}}\mu _{i}\langle \text{%
\textbf{y}}_{i}|t_{\pm k}\rangle |\text{\textbf{y}}_{i}\rangle  \notag \\
& =\sum_{h_{1},...,h_{\mathsf{N}}=1}^{p}\left( \frac{q^{\mp kh_{\mathsf{N}}}%
}{p^{1/2}}\right) ^{\mathtt{e}_{\mathsf{N}}}\prod_{a=1}^{[\mathsf{N}%
]}Q_{t}(\eta _{a}^{(h_{a})})\prod_{1\leq b<a\leq \lbrack \mathsf{N}]}((\eta
_{a}^{(h_{a})})^{2}-(\eta _{b}^{(h_{b})})^{2})\frac{|\eta
_{1}^{(h_{1})},...,\eta _{\mathsf{N}}^{(h_{\mathsf{N}})}\rangle }{%
\prod_{b=1}^{[\mathsf{N}]}\omega _{b}(\eta _{b}^{(h_{b})})},
\label{eigenT-r}
\end{align}%
corresponding to the eigenvalue $t(\lambda )\in \Sigma _{\mathsf{T}}^{k}$,
where the index $k\in \{0,...,(p-1)/2\}$ appears only for $\mathsf{N}$ even
and indicates that $|t_{\pm k}\rangle $ is also a $\Theta $-eigenstate with
eigenvalue $q^{\pm k}$. Here $Q_{t}(\lambda )$ is the solution of the Baxter
equation (\ref{EQ-Baxter-L}) defined in Proposition \ref{SOV-T-Q}. In a
similar way we can prove that, up to an overall normalization, the left $%
\mathsf{T}$-eigenstate has the following SOV-representation:%
\begin{align}
\langle t_{\pm k}|& =\sum_{i=1}^{p^{\mathsf{N}}}\mu _{i}\langle t_{\pm k}|%
\text{\textbf{y}}_{i}\rangle \langle \text{\textbf{y}}_{i}|  \notag \\
& =\sum_{h_{1},...,h_{\mathsf{N}}=1}^{p}\left( \frac{q^{\pm kh_{\mathsf{N}}}%
}{p^{1/2}}\right) ^{\mathtt{e}_{\mathsf{N}}}\prod_{a=1}^{[\mathsf{N}]}\bar{Q}%
_{t}(\eta _{a}^{(h_{a})})\prod_{1\leq b<a\leq \lbrack \mathsf{N}]}(\eta
_{a}^{(h_{a})}-\eta _{b}^{(h_{b})})\frac{\langle \eta
_{1}^{(h_{1})},...,\eta _{\mathsf{N}}^{(h_{\mathsf{N}})}|}{\prod_{b=1}^{[%
\mathsf{N}]}\omega _{b}(\eta _{b}^{(h_{b})})},  \label{eigenT-l}
\end{align}%
where $\bar{Q}_{t}(\lambda )$ is the unique (up to quasi-constants)
polynomial solution of the Baxter functional equation:%
\begin{equation}
t(\lambda )\bar{Q}_{t}(\lambda )=\bar{d}(\lambda )\bar{Q}_{t}(\lambda
q^{-1})+\bar{a}(\lambda )\bar{Q}_{t}(\lambda q).  \label{EQ-Baxter-L}
\end{equation}%
\textbf{Remark 4.} In the gauge fixed by (\ref{L-R-gauge}), this Baxter
equation reads:%
\begin{equation}
t(\lambda )\bar{Q}_{t}(-\lambda )=q^{\mathsf{N}}a(\lambda )\bar{Q}%
_{t}(-\lambda q^{-1})+q^{-\mathsf{N}}d(\lambda )\bar{Q}_{t}(-\lambda q),
\end{equation}%
and so, up to quasi-constants, we have:%
\begin{equation}
\bar{Q}_{t}(\lambda )=\lambda ^{\chi _{\mathsf{N}}}Q_{t}(-\lambda
),\,\,\,\,\,\,\,\,\chi _{\mathsf{N}}=\mathsf{N}\,\,\mathsf{mod}\,p,\text{ }%
0\leq \chi _{\mathsf{N}}\leq p-1.
\end{equation}

\section{Decomposition of the identity in the $\mathsf{T}$-eigenbasis}

Here we use the results of the previous section to compute the action of
covectors on vectors which in the left and right SOV-basis have a \textit{%
separate form} similar to that of the transfer matrix eigenstates.

\subsection{Action of left separate states on right separate states}

Let us start giving the following definition of a left $\langle \alpha _{k}| 
$ and a right $|\beta _{k}\rangle $ separate states; they are respectively
a covector and a vector which have the following \textit{separate form} in
terms of the SOV-decomposition of the identity:%
\begin{align}
\langle \alpha _{k}|& =\sum_{h_{1},...,h_{\mathsf{N}}=1}^{p}\left( \frac{%
q^{kh_{\mathsf{N}}}}{p^{1/2}}\right) ^{\mathtt{e}_{\mathsf{N}}}\prod_{a=1}^{[%
\mathsf{N}]}\alpha _{a}(\eta _{a}^{(h_{a})})\prod_{1\leq b<a\leq \lbrack 
\mathsf{N}]}(\left( \eta _{a}^{(h_{a})}\right) ^{2}-\left( \eta
_{b}^{(h_{b})}\right) ^{2})\frac{\langle \eta _{1}^{(h_{1})},...,\eta _{%
\mathsf{N}}^{(h_{\mathsf{N}})}|}{\prod_{b=1}^{[\mathsf{N}]}\omega _{b}(\eta
_{b}^{(h_{b})})},  \label{Fact-left-SOV} \\
|\beta _{k}\rangle & =\sum_{h_{1},...,h_{\mathsf{N}}=1}^{p}\left( \frac{%
q^{-kh_{\mathsf{N}}}}{p^{1/2}}\right) ^{\mathtt{e}_{\mathsf{N}%
}}\prod_{a=1}^{[\mathsf{N}]}\beta _{a}(\eta _{a}^{(h_{a})})\prod_{1\leq
b<a\leq \lbrack \mathsf{N}]}(\left( \eta _{a}^{(h_{a})}\right) ^{2}-\left(
\eta _{b}^{(h_{b})}\right) ^{2})\frac{|\eta _{1}^{(h_{1})},...,\eta _{%
\mathsf{N}}^{(h_{\mathsf{N}})}\rangle }{\prod_{b=1}^{[\mathsf{N}]}\omega
_{b}(\eta _{b}^{(h_{b})})},  \label{Fact-right-SOV}
\end{align}%
where the index $k$ appears only for $\mathsf{N}$ even. It is easy to see that such states generate the whole space of states of the model (in particular the $\mathsf{T}$-eigenbasis is just of this type). The interest toward
these kind of states is due to the following:

\begin{proposition}
Let $\langle \alpha _{k}|$ and $|\beta _{k'}\rangle $ two separate states of
the form $(\ref{Fact-left-SOV})$ and $(\ref{Fact-right-SOV})$, respectively,
then it holds:%
\begin{equation}
\langle \alpha _{k}|\beta _{k'}\rangle =\delta _{k,k'}^{\mathtt{e}_{\mathsf{N}%
}}\det_{[\mathsf{N}]}||\mathcal{M}_{a,b}^{\left( \alpha ,\beta \right) }||%
\text{ \ \ with \ }\mathcal{M}_{a,b}^{\left( \alpha ,\beta \right) }\equiv
\left( \eta _{a}^{(0)}\right) ^{2(b-1)}\sum_{h=1}^{p}\frac{\alpha _{a}(\eta
_{a}^{(h)})\beta _{a}(\eta _{a}^{(h)})}{\omega _{a}(\eta _{a}^{(h)})}%
q^{2(b-1)h}.
\end{equation}
\end{proposition}

\begin{proof}
From the SOV-decomposition, we have:%
\begin{equation}
\langle \alpha _{k}|\beta _{k'}\rangle =\left( \sum_{h_{\mathsf{N}}=1}^{p}%
\frac{q^{(k-k')h_{\mathsf{N}}}}{p}\right) ^{\mathtt{e}_{\mathsf{N}%
}}\sum_{h_{1},...,h_{[\mathsf{N}]}=1}^{p}V(\left( \eta _{1}^{(h_{1})}\right)
^{2},...,\left( \eta _{\lbrack \mathsf{N}]}^{(h_{[\mathsf{N}]})}\right)
^{2})\prod_{a=1}^{[\mathsf{N}]}\frac{\alpha _{a}(\eta _{a}^{(h_{a})})\beta
_{a}(\eta _{a}^{(h_{a})})}{\omega _{a}(\eta _{a}^{(h_{a})})},
\end{equation}%
where $V(x_{1},...,x_{\mathsf{N}})\equiv \prod_{1\leq b<a\leq \lbrack 
\mathsf{N}]}(x_{a}-x_{b})$ is the Vandermonde determinant. From this formula
by using the multilinearity of the determinant w.r.t. the rows we prove the
proposition.
\end{proof}

In Section \ref{Def-cyclic-rep}, we have associated to the linear space $%
\mathcal{R}_{\mathsf{N}}$ the structure of an Hilbert space by introducing a
scalar product.  Then it is clear that the above determinant formula
represents also the formula for the scalar product of the two vectors $%
\left( \langle \alpha _{k}|\right) ^{\dagger }$  and $|\beta _{h}\rangle $ in 
$\mathcal{R}_{\mathsf{N}}$. It is worth pointing out that the vector $\left(
\langle \alpha _{k}|\right) ^{\dagger }\in \mathcal{R}_{\mathsf{N}}$ is a separate state in the right $\mathsf{C}$-eigenbasis, as it simply
follows from the hermitian conjugation properties of the Yang-Baxter
generators reported in Section \ref{Def-cyclic-rep}. Then these results can
be considered as the SOV analogue of the scalar product formulae \cite%
{Salv89,KitMT99} computed  in the framework of the algebraic
Bethe ansatz. However, we want to stress that the determinant formulae obtained here are not restricted to the case in which one of the two states
is an eigenstate of the transfer matrix, on the contrary to what happens for
the scalar product formulae in the framework of the algebraic Bethe ansatz.
Finally, we can prove directly from these formula the following:

\begin{corollary}
Transfer matrix eigenstates corresponding to different eigenvalues are
orthogonal states.
\end{corollary}

\begin{proof}
Let us denote with $|t_{k}\rangle $ and $|t_{h}^{\prime }\rangle $ two
eigenstates of $\mathsf{T}(\lambda )$ with eigenvalues $t_{k}(\lambda )$ and 
$t_{h}^{\prime }(\lambda )$ for $\mathsf{N}$ odd and with $\Theta $
eigenvalues $q^{k}$ and $q^{h}$ for $\mathsf{N}$ even. To prove the
corollary, we have to prove that:%
\begin{equation}
\det_{\lbrack \mathsf{N}]}||\Phi _{a,b}^{\left( t,t^{\prime }\right) }||=0%
\text{ \ \ with \ }\Phi _{a,b}^{\left( t,t^{\prime }\right) }\equiv \left(
\eta _{a}^{(0)}\right) ^{2(b-1)}\sum_{c=1}^{p}\frac{Q_{t^{\prime }}(\eta
_{a}^{(c)})\bar{Q}_{t}(\eta _{a}^{(c)})}{\omega _{a}(\eta _{a}^{(c)})}%
q^{2(b-1)c},  \label{orth-cond}
\end{equation}%
with $h=k$ for $\mathsf{N}$ even. To prove (\ref{L-R-gauge}), it is enough
to show the existence of a non-zero vector V$^{\left( t,t^{\prime }\right) }$
such that:%
\begin{equation}
\sum_{b=1}^{[\mathsf{N}]}\Phi _{a,b}^{\left( t,t^{\prime }\right) }\text{V}%
_{b}^{\left( t,t^{\prime }\right) }=0\text{ \ \ \ \ }\forall a\in \{1,...,[%
\mathsf{N}]\}.  \label{zero-eigenvector}
\end{equation}%
For simplicity, we construct this vector in the gauge (\ref{L-R-gauge})
where it results:%
\begin{equation}
\omega _{a}(\eta _{a}^{(h)})=\left( \eta _{a}^{(h)}\right) ^{[\mathsf{N}]-1}.
\end{equation}%
Let us recall that the eigenvalues of the transfer matrix are even Laurent
polynomials of degree $\bar{\mathsf{N}}$ of the form:%
\begin{eqnarray}
t_{h}(\lambda ) &=&\mathtt{e}_{\mathsf{N}}\left( \prod_{a=1}^{\mathsf{N}}%
\frac{\kappa _{a}\xi _{a}^{\pm 1}}{i}\right) (q^{h}+q^{-h})(\lambda ^{%
\mathsf{N}}+\lambda ^{-\mathsf{N}})+\sum_{b=1}^{[\mathsf{N}]}c_{b}\lambda
^{-[\mathsf{N}]-1+2b}, \\
t_{h}^{\prime }(\lambda ) &=&\mathtt{e}_{\mathsf{N}}\left( \prod_{a=1}^{%
\mathsf{N}}\frac{\kappa _{a}\xi _{a}^{\pm 1}}{i}\right)
(q^{h}+q^{-h})(\lambda ^{\mathsf{N}}+\lambda ^{-\mathsf{N}})+\sum_{b=1}^{[%
\mathsf{N}]}c_{b}^{\prime }\lambda ^{-[\mathsf{N}]-1+2b},
\end{eqnarray}%
so if we define:%
\begin{equation}
\text{V}_{b}^{\left( t,t^{\prime }\right) }\equiv c_{b}^{\prime }-c_{b}\text{%
\ \ \ }\forall b\in \{1,...,[\mathsf{N}]\},
\end{equation}%
it results:%
\begin{equation}
\sum_{b=1}^{[\mathsf{N}]}\Phi _{a,b}^{\left( t,t^{\prime }\right) }\text{V}%
_{b}^{\left( t,t^{\prime }\right) }=\sum_{c=1}^{p}Q_{t^{\prime }}(\eta
_{a}^{(c)})\bar{Q}_{t}(\eta _{a}^{(c)})(t_{h}^{\prime }(\eta
_{a}^{(c)})-t_{h}(\eta _{a}^{(c)})).  \label{zero-eigenvector-1}
\end{equation}%
We can use now the Baxter equations (\ref{EQ-Baxter-R}) and (\ref%
{EQ-Baxter-L}), with the chosen gauge (\ref{L-R-gauge}), to rewrite:%
\begin{align}
Q_{t^{\prime }}(\eta _{a}^{(h_{a})})\bar{Q}_{t}(\eta
_{a}^{(h_{a})})(t^{\prime }(\eta _{a}^{(h_{a})})-t(\eta _{a}^{(h_{a})}))&
=a(\eta _{a}^{(h_{a})})Q_{t^{\prime }}(\eta _{a}^{(h_{a}-1)})\bar{Q}%
_{t}(\eta _{a}^{(h_{a})})\notag \\
&+d(\eta _{a}^{(h_{a})})Q_{t^{\prime }}(\eta
_{a}^{(h_{a}+1)})\bar{Q}_{t}(\eta _{a}^{(h_{a})})  \notag \\
& -d(\eta _{a}^{(h_{a}-1)})Q_{t^{\prime }}(\eta _{a}^{(h_{a})})\bar{Q}%
_{t}(\eta _{a}^{(h_{a}-1)})\notag \\
&-a(\eta _{a}^{(h_{a}+1)})Q_{t^{\prime }}(\eta
_{a}^{(h_{a})})\bar{Q}_{t}(\eta _{a}^{(h_{a}+1)}),
\end{align}%
and by substituting it in (\ref{zero-eigenvector-1}) we get (\ref%
{zero-eigenvector}).
\end{proof}

\subsection{Decomposition of the identity in the \textsf{T}-eigenbasis}

Let us remark that the diagonalizability and simplicity of the transfer
matrix spectrum implies the following decomposition of the identity in
the left and right $\mathsf{T}$-eigenbasis:%
\begin{equation}
\mathbb{I=}\sum_{k=0}^{\mathtt{e}_{\mathsf{N}}\left( p-1\right)
/2}\sum_{t_{k}(\lambda )\in \Sigma _{\mathsf{T}}^{k}}\frac{|t_{k}\rangle
\langle t_{k}|}{\langle t_{k}|t_{k}\rangle },  \label{Id-decomp}
\end{equation}%
where%
\begin{equation}
\langle t_{k}|t_{k}\rangle =\det_{[\mathsf{N}]}||\Phi _{a,b}^{\left(
t_{k},t_{k}\right) }||\text{ \ with }\Phi _{a,b}^{\left( t_{k},t_{k}\right)
}\equiv (\eta _{a}^{(0)})^{2(b-1)}\sum_{c=1}^{p}\frac{Q_{t}(\eta _{a}^{(c)})%
\bar{Q}_{t}(\eta _{a}^{(c)})}{\omega _{a}(\eta _{a}^{(c)})}q^{2(b-1)c},
\end{equation}%
is the action of the covector $\langle t_{k}|$ on the vector $|t_{k}\rangle $
as defined in Section \ref{SOV-T-eigenstates}. It is worth to note that in
the representations which defines a normal transfer matrix $\mathsf{T}%
(\lambda )$, the generic covector \underline{$\langle t_{k}|$}$\equiv \left(
|t_{k}\rangle \right) ^{\dagger }$, dual to the right $\mathsf{T}$%
-eigenstate $|t_{k}\rangle $, is itself a left eigenstate of $\mathsf{T}%
\left( \lambda \right) $ which taking into account  the simplicity of the $\mathsf{T}$%
-spectrum has to satisfy the following identity \underline{$\langle t_{k}|$}$%
\equiv \alpha _{t_{k}}\langle t_{k}|$, where $\langle t_{k}|$ is the left $%
\mathsf{T}$-eigenstate defined in (\ref{eigenT-l}). Of course, in these
representations the following identities hold: 
\begin{equation}
\frac{|t_{k}\rangle \langle t_{k}|}{\langle t_{k}|t_{k}\rangle }=\frac{%
|t_{k}\rangle \underline{\langle t_{k}|}}{\left\Vert |t_{k}\rangle
\right\Vert ^{2}}
\end{equation}%
where $\left\Vert |t_{k}\rangle \right\Vert $ is the positive norm of the
eigenvector $|t_{k}\rangle $ in the Hilbert space $\mathcal{R}_{\mathsf{N}}$. The above discussion implies the relevance of computing explicitly the norm
of the transfer matrix eigenstates (\ref{eigenT-r}) as it allows to fix the relative normalization $\alpha _{t_{k}}$ thanks to the identity $\alpha_{t_{k}}=\left\Vert |t_{k}\rangle \right\Vert ^{2}/\langle t_{k}|t_{k}\rangle $ and then it allows to take these left and right states as the one being the exact dual of the other. This interesting issue is
currently under analysis.

\section{SOV-representation of local operators}

The determination of the scalar product formulae, presented in the previous
section, is the first main step to compute matrix elements of local
operators. The second one is to get the reconstruction of local operators in
terms of the generators of the Yang-Baxter algebra, i.e. to invert the map
which from the local operators in the Lax matrices leads to the monodromy
matrix elements. Indeed, the solution of such an inverse problem allows to
compute the action of local operators on the eigenstates of the transfer
matrix. Together with the scalar product formulae it leads to the
determination of the matrix elements of local operators.

The first reconstruction of local operators has been achieved in \cite%
{KitMT99}, in the case of the XXZ spin 1/2 chain. In \cite{MaiT00}, it has
been extended to fundamental lattice models, i.e. those with isomorphic\
auxiliary and local quantum space, for which the monodromy matrix becomes
the permutation operator at a special value of the spectral parameter. The
reconstruction also applies to non-fundamental lattice models, as it was
shown in \cite{MaiT00} for the higher spin XXX chains by using the fusion
procedure \cite{KulRS81}. In the case of the sine-Gordon model this type of
reconstruction is still missing and the only known results are those given
by T. Oota based on the use of quantum projectors \cite{Oota03}. However, it
is worth recalling that Oota's results only lead to the reconstruction of
some local operators of the lattice sine-Gordon model. In this section, we
will show how to obtain all the local operators of the sine-Gordon model for
the cyclic representations which occur at rational values of the coupling
constant $\beta ^{2}$.

\subsection{Oota's reconstruction of a class of local operators}

Here we recall some of the results of Oota \cite{Oota03}\ which lead to the
reconstruction of a certain class of local operators in the sine-Gordon
model.

The Lax operator $\mathsf{L}_{n}(\lambda )$ has the following factorization
in terms of quantum projectors:%
\begin{align}
\mathsf{L}_{n}(\mu _{n,+})& =P_{n,+}Q_{n,+}\equiv \kappa _{n}\left( 
\begin{array}{c}
\mathsf{u}_{n}^{1/2}\left( \mathsf{v}_{n}\kappa _{n}+\mathsf{v}%
_{n}^{-1}\kappa _{n}^{-1}\right) \\ 
\mathsf{u}_{n}^{-1/2}\left( \mathsf{v}_{n}\kappa _{n}^{-1}+\mathsf{v}%
_{n}^{-1}\kappa _{n}\right)%
\end{array}%
\right) \left( 
\begin{array}{cc}
\mathsf{u}_{n}^{1/2} & \mathsf{u}_{n}^{-1/2}%
\end{array}%
\right) ,  \label{L-f+} \\
\mathsf{L}_{n}(\mu _{n,-})& =P_{n,-}Q_{n,-}\equiv \kappa _{n}\left( 
\begin{array}{c}
\mathsf{u}_{n}^{1/2} \\ 
\mathsf{u}_{n}^{-1/2}%
\end{array}%
\right) \left( 
\begin{array}{cc}
\left( \mathsf{v}_{n}\kappa _{n}+\mathsf{v}_{n}^{-1}\kappa _{n}^{-1}\right) 
\mathsf{u}_{n}^{1/2} & \left( \mathsf{v}_{n}\kappa _{n}^{-1}+\mathsf{v}%
_{n}^{-1}\kappa _{n}\right) \mathsf{u}_{n}^{-1/2}%
\end{array}%
\right) ,  \label{L-f-}
\end{align}%
when computed in the zeros $\mu _{n,\pm }$ of the quantum determinant; such
factorization properties are at the basis of the following Oota's
reconstruction.

\begin{proposition}
The local operators $\mathsf{u}_{n}$ and $\alpha _{0,n}\equiv \left( (q^{-1}%
\mathsf{v}_{n}^{2}+\kappa _{n}^{2})/\left( q^{-1}\mathsf{v}_{n}^{2}\kappa
_{n}^{2}+1\right) \right) \mathsf{u}_{n}^{-1}$ admit the reconstructions:%
\begin{eqnarray}
\mathsf{u}_{n} &=&\mathsf{U}_{n}\mathsf{B}^{-1}(\mu _{n,+})\mathsf{A}(\mu
_{n,+})\mathsf{U}_{n}^{-1}=\mathsf{U}_{n}\mathsf{D}^{-1}(\mu _{n,+})\mathsf{C%
}(\mu _{n,+})\mathsf{U}_{n}^{-1},  \label{IPS-1} \\
&&  \notag \\
\alpha _{0,n} &=&\mathsf{U}_{n}\mathsf{A}^{-1}(\mu _{n,-})\mathsf{B}(\mu
_{n,-})\mathsf{U}_{n}^{-1}=\mathsf{U}_{n}\mathsf{C}^{-1}(\mu _{n,-})\mathsf{D%
}(\mu _{n,-})\mathsf{U}_{n}^{-1},  \label{IPS-2}
\end{eqnarray}%
where the shift operator $\mathsf{U}_{n}$ brings the quantum sites from $1$
to $n-1$ to the right end of the chain:%
\begin{equation}
\mathsf{U}_{n}\mathsf{M}_{1,...,\mathsf{N}}(\lambda )\mathsf{U}%
_{n}^{-1}\equiv \mathsf{M}_{n,...,\mathsf{N},1,...,n-1}(\lambda )\equiv 
\mathsf{L}_{n-1}(\lambda )\cdots \mathsf{L}_{1}(\lambda )\mathsf{L}_{\mathsf{%
N}}(\lambda )\cdots \mathsf{L}_{n}(\lambda ).  \label{Def-Un}
\end{equation}
\end{proposition}

It is then clear that the formulae (\ref{IPS-1})-(\ref{IPS-2}) allow to
reconstruct all the powers $\mathsf{u}_{n}^{k}=\mathsf{U}_{n}\left( \mathsf{B%
}^{-1}(\mu _{n,+})\mathsf{A}(\mu _{n,+})\right) ^{k}\mathsf{U}_{n}^{-1}$ of
the local operators $\mathsf{u}_{n}$ but they do not give a direct
reconstruction of the local operators $\mathsf{v}_{n}$; indeed, only
rational functions like $(q^{-1}\mathsf{v}_{n}^{2}+\kappa _{n}^{2})/\left(
q^{-1}\mathsf{v}_{n}^{2}\kappa _{n}^{2}+1\right) $ are obtained.

Let us make some comments on the shift operators $\mathsf{U}_{n}$. The
definition (\ref{Def-Un}) characterizes the shift operator \textsf{U}$_{n}$
up to a constant and implies, by the cyclicity invariance of the trace,
their commutativity with the transfer matrix $\mathsf{T}(\lambda )$.
Moreover, in the case of even chain, the shift operators $\mathsf{U}_{n}$
clearly commute also with the $\Theta $-charge. Then, in the cyclic
representations of the sine-Gordon model under consideration, the simplicity
of the transfer matrix spectrum implies:%
\begin{equation}
\mathsf{U}_{n}|t_{k}\rangle =\varphi _{n}^{(t_{k})}|t_{k}\rangle ,
\end{equation}%
where $|t_{k}\rangle $ is the generic eigenstate of $\mathsf{T}(\lambda )$
for odd chain and of ($\mathsf{T}(\lambda )$,$\Theta $) for the even chain.
In particular, this implies that the shift operators only produce a
prefactor in the form factors of local operators which is one if left and
right eigenstates are dual of each other\footnote{%
Translational invariance for the limit of the homogeneous chain.}. It is
worth remarking that in\footnote{%
This result is there attributed to V. Korepin, private communications.} \cite%
{TarTF83}, for the special case of highest weight representations of the
even lattice sine-Gordon model, it has been shown that:%
\begin{equation}
\mathsf{U}_{n}|t\rangle \propto \prod_{a=1}^{n-1}t(\mu _{a})|t\rangle
\propto \prod_{a=1}^{n-1}\frac{Q_{t}(\mu _{a}/q)}{Q_{t}(\mu _{a})}|t\rangle ,
\label{Un-eigenvalues}
\end{equation}%
where $\mu _{a}$ are zeros of the quantum determinant in these
representations. This result is interesting as it shows that the shift
operators \textsf{U}$_{n}$ for non-fundamental lattice models, like the
sine-Gordon model, are characterized by the same type of eigenvalues they
have in fundamental lattice models, like the XXZ spin 1/2 chain. However,
the proof of (\ref{Un-eigenvalues}) presented in \cite{TarTF83} is
representation dependent, as it is based on the algebraic Bethe ansatz
representation of the transfer matrix eigenstates. Then, an independent
proof is required for cyclic representations of the sine-Gordon model and it
will be given directly in a future publication \cite{GroMN12} in the more
general cyclic representations of the 6-vertex Yang-Baxter algebra
associated to the $\tau_2$-model.

\subsection{Inverse problem solution for all local operators}

Here, we show how to complete the reconstruction of local operators by
solving the inverse problem for the local operators $\mathsf{v}_{n}$ and
their powers. The main ingredient used will be the cyclicity of the
representations of the sine-Gordon model here analyzed.

Let us define the local operators:%
\begin{equation}
\beta _{k,n}\equiv \mathsf{u}_{n}^{k}\alpha _{0,n}\mathsf{u}_{n}^{1-k}=\frac{%
q^{2k-1}\mathsf{v}_{n}^{2}+\kappa _{n}^{2}}{q^{2k-1}\mathsf{v}_{n}^{2}\kappa
_{n}^{2}+1},
\end{equation}%
then the following proposition holds:

\begin{proposition}
\label{IPS}In the cyclic representations of the sine-Gordon model, the local
operators $\mathsf{v}_{n}^{2k}$ admit the following reconstruction:%
\begin{equation}
\mathsf{v}_{n}^{2k}=\frac{\left( -1\right) ^{k}(v_{n}^{2p}\kappa _{n}^{2p}+1)%
}{p\kappa _{n}^{2k}(\kappa _{n}^{2}-\kappa _{n}^{-2})}%
\sum_{a=0}^{p-1}q^{-k(2a-1)}\beta _{a,n},\text{ \ \ \ for }k\in
\{1,...,p-1\},  \label{IPS-3}
\end{equation}%
where:%
\begin{equation}
\beta _{k,n}=\mathsf{U}_{n}\left[ \left( \mathsf{B}^{-1}(\mu _{n,+})\mathsf{A%
}(\mu _{n,+})\right) ^{k}\mathsf{A}^{-1}(\mu _{n,-})\mathsf{B}(\mu
_{n,-})\left( \mathsf{B}^{-1}(\mu _{n,+})\mathsf{A}(\mu _{n,+})\right) ^{1-k}%
\right] \mathsf{U}_{n}^{-1}.  \label{IPS-4}
\end{equation}
\end{proposition}

\begin{proof}
In our cyclic representations the local operators $\mathsf{u}_{n}$ and $%
\mathsf{v}_{n}$ satisfy the property that $\mathsf{u}_{n}^{p}$ and $\mathsf{v%
}_{n}^{p}$ are central, i.e. $\mathsf{u}_{n}^{p}$ and $\mathsf{v}_{n}^{p}$
are just numbers $u_{n}^{p}$ and $v_{n}^{p}$ which characterize our
representations. Then the following identity holds:%
\begin{equation}
\prod_{j=0}^{p-1}(q^{2j-1}\mathsf{v}_{n}^{2}\kappa
_{n}^{2}+1)=1+v_{n}^{2p}\kappa _{n}^{2p},
\end{equation}%
and so:%
\begin{equation}
\frac{v_{n}^{2p}\kappa _{n}^{2p}+1}{q^{2k-1}\mathsf{v}_{n}^{2}\kappa
_{n}^{2}+1}=\sum_{a=0}^{p-1}(-1)^{a}q^{a(2k-1)}\mathsf{v}_{n}^{2a}\kappa
_{n}^{2a}.
\end{equation}%
The previous formula allows to rewrite the rational function $\beta _{k,n}$
as a finite sum in powers of $\mathsf{v}_{n}^{2}$:%
\begin{equation}
\beta _{k,n}=\frac{v_{n}^{2p}\kappa _{n}^{2(p-1)}+\kappa _{n}^{2}+(\kappa
_{n}^{2}-\kappa _{n}^{-2})\sum_{a=1}^{p-1}(-1)^{a}q^{a(2k-1)}\mathsf{v}%
_{n}^{2a}\kappa _{n}^{2a}}{v_{n}^{2p}\kappa _{n}^{2p}+1},
\end{equation}%
then by taking the discrete Fourier transformation, we get the
reconstructions (\ref{IPS-3}), plus the sum rules:%
\begin{equation}
\sum_{a=0}^{p-1}\beta _{a,n}=\frac{p v_{n}^{2p}\kappa _{n}^{2(p-1)}+\kappa
_{n}^{2}}{v_{n}^{2p}\kappa _{n}^{2p}+1}.
\end{equation}%
Finally, the representation (\ref{IPS-4}) for the $\beta _{a,n}$ are
trivially derived by (\ref{IPS-1})-(\ref{IPS-2}).
\end{proof}

Note that thanks to the identities $\mathsf{v}_{n}^{k}=\mathsf{v}%
_{n}^{2h}/v_{n}^{p}$ for $k=2h-p$ odd integer smaller than $p$, the formulae
(\ref{IPS-3}) indeed lead to the reconstruction of all the powers $\mathsf{v}%
_{n}^{k}$ for $k\in \{1,...,p-1\}$. Then the previous proposition together
with Oota's reconstructions leads to the announced complete reconstruction
of local operators for cyclic representations of the sine-Gordon model.

\subsection{SOV-representations of all local operators}

In order to compute the action of the local operators $\mathsf{v}_{n}^{k}$
and $\mathsf{u}_{n}^{k}$ on eigenstates of the transfer matrix and
eventually obtain their form factors, it is necessary to first derive their
SOV-representations\footnote{%
To simply the notations we chose to present the results in this subsection
only for the case $\mathsf{N}$ odd and for the representations with $\mathsf{%
v}_{n}^{p}=\mathsf{u}_{n}^{p}=1.$}. Here, we show how these
SOV-representations can be obtained from the previously given solutions of
the inverse problem. In order to simplify the presentation, we introduce
here explicitly the operators\footnote{%
To simplify the exposition, we have decided to keep as simple as possible
the notations for these operators, we hope that nevertheless the difference
with the corresponding eigenvalues is clear.} $\eta _{1},...,\eta _{\mathsf{N%
}},\eta _{\mathsf{A}}$ and $\eta _{\mathsf{D}}$ defined by the following
actions on the generic element $\langle \mathbf{k}|\equiv \langle \eta
_{1}^{\left( k_{1}\right) },...,\eta _{\mathsf{N}}^{\left( k_{\mathsf{N}%
}\right) }|$ of the left $\mathsf{B}$-eigenbasis:%
\begin{equation}
\langle \mathbf{k}|\eta _{a}=\eta _{a}^{\left( k_{a}\right) }\langle \mathbf{%
k}|,\text{ \ }\forall a\in \{1,...,\mathsf{N}\}\text{, \ \ }\langle \mathbf{k%
}|\eta _{\mathsf{A}}=\eta _{\mathbf{k,}\mathsf{A}}\langle \mathbf{k}|\text{
\ and \ \ }\langle \mathbf{k}|\eta _{\mathsf{D}}=\eta _{\mathbf{k,}\mathsf{D}%
}\langle \mathbf{k}|,
\end{equation}%
where the corresponding eigenvalues $\eta _{a}^{\left( k_{a}\right) }$, $%
\eta _{\mathbf{k,}\mathsf{A}}$ and $\eta _{\mathbf{k,}\mathsf{D}}$ are the
complex numbers defined in Section \ref{SOV-Left}. The following lemma is
important as it solves the combinatorial problem related to the computations
of the SOV-representations of monomials in the Yang-Baxter generators:

\begin{lemma}
The operator $\left( \mathsf{B}^{-1}(\lambda )\mathsf{A}(\lambda )\right)
^{k}$ has the following left SOV-representation:%
\begin{align}
\left( \mathsf{B}^{-1}(\lambda )\mathsf{A}(\lambda )\right) ^{k}& =\text{%
\textsc{k}}^{-k}\sum_{\substack{ \bar{\alpha}\equiv \{\alpha _{1},...,\alpha
_{\mathsf{N}}\}\in \mathbb{N}_{0}^{\mathsf{N}}:  \\ \sum_{h=1}^{\mathsf{N}%
}\alpha _{h}=k}}\left[ 
\begin{array}{c}
k \\ 
\bar{\alpha}%
\end{array}%
\right] \prod_{j=1}^{\mathsf{N}}\left( \prod_{h=0}^{\alpha _{j}-1}\frac{%
a(\eta _{j}q^{-h})}{(\lambda q^{h}/\eta _{j}-\eta _{j}/\lambda q^{h})}\right.
\notag \\
& \times \left. \prod_{i\neq j,i=1}^{\mathsf{N}}\prod_{h=\alpha _{i}-\alpha
_{j}+1}^{\alpha _{i}}\frac{1}{(\eta _{j}q^{h}/\eta _{i}-\eta _{i}/\eta
_{j}q^{h})}\right) \prod_{j=1}^{\mathsf{N}}\mathsf{T}_{j}^{-\alpha _{j}},
\label{A/B^k}
\end{align}%
acting on the state $\langle \eta _{1},...,\eta _{\mathsf{N}}|$, where:%
\begin{equation}
\left[ 
\begin{array}{c}
k \\ 
\bar{\alpha}%
\end{array}%
\right] \equiv \frac{\lbrack k]!}{\prod_{j=1}^{\mathsf{N}}\left[ \alpha _{j}%
\right] !},\text{ }[k]!\equiv \lbrack k][k-1]\cdots \lbrack 1],\text{ }%
[a]\equiv \frac{q^{a}-q^{-a}}{q-q^{-1}}.
\end{equation}
\end{lemma}

\begin{proof}
Here we use the commutation relations:%
\begin{equation}
\mathsf{T}_{a}^{\pm }\eta _{b}=q^{\pm \delta _{ab}}\eta _{b}\mathsf{T}%
_{a}^{\pm },
\end{equation}%
and the following left SOV-representation:%
\begin{equation}
\mathsf{B}^{-1}(\lambda )\mathsf{A}(\lambda )=\text{\textsc{k}}%
^{-1}\sum_{a=1}^{\mathsf{N}}\frac{a(\eta _{a})}{({\lambda }/{\eta _{a}}-{%
\eta _{a}}/{\lambda })}\prod_{b\neq a}\frac{1}{({\eta _{a}}/{\eta _{b}}-{%
\eta _{b}}/{\eta _{a}})}\mathsf{T}_{a}^{-1},
\end{equation}%
then the lemma holds for $k=1$ and we prove it by induction for $k>1$. Let
us take $\mathsf{N}$ integers $\alpha _{i}$: 
\begin{equation}
\sum_{a=1}^{\mathsf{N}}\alpha _{i}=k,
\end{equation}%
from which we define the set of integers $I=\{i\in \{1,...,\mathsf{N}%
\}:\alpha _{i}\neq 0\}$ and $C_{{\bar{\alpha}}}^{(k)}$\ as the coefficient
of $\prod \mathsf{T}_{i}^{-\alpha _{i}}$ in the expansion of the $k$-th
power of $\mathsf{B}^{-1}(\lambda )\mathsf{A}(\lambda )$. By writing $(%
\mathsf{B}^{-1}(\lambda )\mathsf{A}(\lambda ))^{k}=(\mathsf{B}^{-1}(\lambda )%
\mathsf{A}(\lambda ))^{k-1}\mathsf{B}^{-1}(\lambda )\mathsf{A}(\lambda )$
and by using the induction hypothesis for the power $k-1$ of $\mathsf{B}%
^{-1}(\lambda )\mathsf{A}(\lambda )$, we have:%
\begin{align}
C_{{\bar{\alpha}}}^{(k)}& =\text{\textsc{k}}^{-k}\sum_{a\in I}{\left[ 
\begin{array}{c}
{k-1} \\ 
{\bar{\alpha}-\bar{\delta}_{a}}%
\end{array}%
\right] }  \notag \\
& \prod_{j=1}^{N}\prod_{h=0}^{\alpha _{j}-\delta _{a,j}-1}\left( \frac{%
a(\eta _{j}q^{-h})}{({\lambda q^{h}}/{\eta _{j}}-{\eta _{j}}/{\lambda q^{h}})%
}\times \prod_{i\neq j,i=1}^{N}\frac{1}{{q^{\alpha _{i}-\delta _{a,i}-h}\eta
_{j}}/{\eta _{i}}-{\eta _{i}}/{q^{\alpha _{i}-\delta _{a,i}-h}\eta _{j}}}%
\right)  \notag \\
& \times \frac{a(\eta _{a}q^{-\alpha _{a}+1})}{({\lambda q^{\alpha _{a}-1}}/{%
\eta _{a}}-{\eta _{a}}/{\lambda q^{\alpha _{a}-1}})}\prod_{i\in I\backslash
\{a\}}\frac{1}{{q^{\alpha _{a}-\alpha _{i}-1}\eta _{i}}/{\eta _{a}}-{\eta
_{a}}/{q^{\alpha _{a}-\alpha _{i}-1}\eta _{i}}},
\end{align}%
with $\bar{\delta}_{a}\equiv (\delta _{1,a},\dots ,\delta _{\mathsf{N},a})$.
The first term in the r.h.s. is the coefficient of $\prod \mathsf{T}%
_{i}^{-\alpha _{i}+\delta _{a,i}}$ in $(\mathsf{B}^{-1}(\lambda )\mathsf{A}%
(\lambda ))^{k-1}$ and the second is the coefficient of $\mathsf{T}_{a}^{-1}$
in $\mathsf{B}^{-1}(\lambda )\mathsf{A}(\lambda )$ once the commutations
between $\prod \mathsf{T}_{i}^{-\alpha _{i}+\delta _{a,i}}$ and the $\eta
_{i}$ have been performed. This can be rewritten as:%
\begin{align}
C_{{\bar{\alpha}}}^{(k)}=\text{\textsc{k}}^{-k}\frac{[k-1]!}{\prod [\alpha
_{i}]!}& \left( \prod_{j=1}^{N}\prod_{h=0}^{\alpha _{j}-1}(\prod_{i\neq
j,i=1}^{N}\frac{1}{{q^{\alpha _{i}-h}\eta _{j}}/{\eta _{i}}-{\eta _{i}}/{%
q^{\alpha _{i}-h}\eta _{j}}})\frac{a(q^{-h}\eta _{j})}{({\lambda q^{h}}/{%
\eta _{j}}-{\eta _{j}}/{\lambda q^{h}})}\right)  \notag \\
& \times \sum_{a\in I}([\alpha _{a}]\prod_{i\in I\backslash \{a\}}\frac{{%
q^{\alpha _{a}}\eta _{i}}/{\eta _{a}}-{\eta _{a}}/{q^{\alpha _{a}}\eta _{i}}%
}{{q^{\alpha _{a}-\alpha _{i}}\eta _{i}}/{\eta _{a}}-{\eta _{a}}/{q^{\alpha
_{a}-\alpha _{i}}\eta _{i}}}),
\end{align}%
which leads to our result when we use the relation:%
\begin{equation}
\sum_{a=1}^{n}[\alpha _{a}]\prod_{i\neq a}\frac{{q^{\alpha _{a}}\eta _{i}}/{%
\eta _{a}}-{\eta _{a}}/{q^{\alpha _{a}}\eta _{i}}}{{q^{\alpha _{a}-\alpha
_{i}}\eta _{i}}/{\eta _{a}}-{\eta _{a}}/{q^{\alpha _{a}-\alpha _{i}}\eta _{i}%
}}=\left[ \sum_{a=1}^{n}\alpha _{a}\right] .
\end{equation}%
The above formula holds for any $n$, for any set of numbers $\eta _{i}$ and
for any non-negative integers $\alpha _{i}$, all of them being in generic position. This is shown by studying the
contour integral and the residues of the function: 
\begin{equation}
g(z)=\frac{1}{z}\prod_{i=1}^n \frac{z-\eta _{i}^{2}}{z-q^{-2\alpha _{i}}\eta _{i}^{2}%
}.
\end{equation}
\end{proof}

\textbf{Remark 5.} Let us point out that the power $p$ of $\mathsf{B}%
^{-1}(\lambda )\mathsf{A}(\lambda )$\ is a central element of the
Yang-Baxter algebra and it reads:%
\begin{equation}
(\mathsf{B}^{-1}(\lambda )\mathsf{A}(\lambda ))^{p}=\mathcal{B}(\Lambda
)^{-1}\mathcal{A}(\Lambda ),  \label{A/B^p-a}
\end{equation}%
as it simply follows from the commutations relations:%
\begin{equation}
\mathsf{B}^{-1}(q\lambda )\mathsf{A}(q\lambda )=\mathsf{A}(\lambda )\mathsf{B%
}^{-1}(\lambda ).
\end{equation}%
Then, it is important to verify that the same result follows from the
previous lemma for $k=p$. In order to prove it, it is enough to use the
following properties of the quantum binomials:%
\begin{equation}
\text{ }{\left[ 
\begin{array}{c}
p \\ 
{\bar{\alpha}}%
\end{array}%
\right] }=\left\{ 
\begin{array}{l}
1\text{ \ \ if }\exists i\in \{1,...,\mathsf{N}\}:\alpha _{i}=p\delta _{a,i}%
\text{\ }\forall a\in \{1,...,\mathsf{N}\}, \\ 
0\text{ \ \ otherwise,}%
\end{array}%
\right.
\end{equation}%
from which the formula (\ref{A/B^k})\ for $k=p$ reads: 
\begin{equation}
(\mathsf{B}^{-1}(\lambda )\mathsf{A}(\lambda ))^{p}=\text{\textsc{k}}%
^{-p}\sum_{a=1}^{p}\frac{\prod_{k=1}^{p}a(q^{k}\eta _{a})}{({\Lambda }/{Z_{a}%
}-{Z_{a}}/{\Lambda })}\prod_{b\neq a}\frac{1}{({Z_{b}}/{Z_{a}}-{Z_{a}}/{Z_{b}%
})},  \label{A/B^p-b}
\end{equation}
and to observe that the r.h.s. of (%
\ref{A/B^p-b}) indeed coincides with $\mathcal{B}(\Lambda )^{-1}\mathcal{A}%
(\Lambda )$.\smallskip

Note that the above lemma gives directly the SOV-representations of the
local operators $\mathsf{u}_{n}^{k}$ and $\alpha _{0,n}^{-k}$ with $k\in
\{1,...,p-1\}$ when we fix the parameter $\lambda $ to $\mu _{n,\varepsilon
} $ with $\varepsilon =+$ and $\varepsilon =-$, respectively. Moreover, it
allows to derive also the SOV-representations of the local operators $%
\mathsf{v}_{n}^{k}$ as it follows:

\begin{corollary}
The local operators $\mathsf{v}_{n}^{2k}$ with $k\in \{1,...,p-1\}$ have the
following left SOV-representation:%
\begin{align}
\mathsf{U}_{n}^{-1}\mathsf{v}_{n}^{2k}\mathsf{U}_{n}& =\text{\textsc{v}}%
_{n}^{(2k)}+\sum_{a=1}^{\mathsf{N}}\sum_{\substack{ \bar{\alpha}\equiv
\{\alpha _{1},...,\alpha _{\mathsf{N}}\}\in \mathbb{\mathsf{N}}_{0}^{\mathsf{%
N}}:  \\ \sum_{h=1}^{\mathsf{N}}\alpha _{h}=p-1}}\left[ 
\begin{array}{c}
p-1 \\ 
\bar{\alpha}%
\end{array}%
\right] \prod_{j=1}^{\mathsf{N}}\left( \prod_{h=0}^{\alpha _{j}-1}\frac{%
a(\eta _{j}q^{-h})}{(\mu _{n,-}q^{h}/\eta _{j}-\eta _{j}/\mu _{n,-}q^{h})}%
\right.  \notag \\
& \times \left. \prod_{i\neq j,i=1}^{\mathsf{N}}\prod_{h=\alpha _{i}-(\alpha
_{j}+\delta _{j,a})+1}^{\alpha _{i}}\frac{1}{(\eta _{j}q^{h}/\eta _{i}-\eta
_{i}/\eta _{j}q^{h})}\right) \text{\textsc{v}}_{n,(a,\bar{\alpha}%
)}^{(2k)}\prod_{j=1}^{\mathsf{N}}\text{T}_{j}^{-(\alpha _{j}+\delta _{j,a})},
\end{align}%
where:%
\begin{align}
\text{\textsc{v}}_{n}^{(2k)}& \equiv \frac{\left( -1\right) ^{k}(\kappa
_{n}^{2p}+1)}{p\kappa _{n}^{2k}(\kappa _{n}^{2}-\kappa _{n}^{-2})}%
\sum_{r=1}^{p}\frac{q^{-k(2r-1)}(q^{r}-q^{-r})}{q^{r}\kappa
_{n}^{2}-q^{-r}\kappa _{n}^{-2}}, \\
&  \notag \\
\text{\textsc{v}}_{n,(a,\bar{\alpha})}^{(2k)}& \equiv \frac{\left( -1\right)
^{k}(\kappa _{n}^{2p}+1)}{p\kappa _{n}^{2k}(\kappa _{n}^{2}-\kappa _{n}^{-2})%
}\sum_{r=1}^{p}\frac{q^{-k(2r-1)}(\kappa _{n}^{2}-\kappa _{n}^{-2})a(\eta
_{a}q^{-\alpha _{a}})}{\left( q^{r}\kappa _{n}^{2}-q^{-r}\kappa
_{n}^{-2}\right) (\mu _{n,+}q^{\alpha _{a}+r}/\eta _{a}-\eta _{a}/\mu
_{n,+}q^{\alpha _{a}+r})}  \notag \\
& \times \prod_{j=1}^{\mathsf{N}}\prod_{h=0}^{r-1}\frac{(\mu _{n,+}q^{\alpha
_{j}+h}/\eta _{j}-\eta _{j}/\mu _{n,+}q^{\alpha _{j}+h})}{(\mu
_{n,+}q^{h}/\eta _{j}-\eta _{j}/\mu _{n,+}q^{h})}.
\end{align}
\end{corollary}

\begin{proof}
This is a consequence of the previous lemma and of the identities:%
\begin{equation}
\mathsf{U}_{n}^{-1}\beta _{k,n}\mathsf{U}_{n}=\frac{\kappa _{n}^{2}-\kappa
_{n}^{-2}}{q^{k}\kappa _{n}^{2}-q^{-k}\kappa _{n}^{-2}}\gamma _{k,n}+\frac{%
q^{k}-q^{-k}}{q^{k}\kappa _{n}^{2}-q^{-k}\kappa _{n}^{-2}},\text{ \ for }%
k\in \{1,...,p-1\},
\end{equation}%
where:%
\begin{equation}
\gamma _{k,n}=\mathsf{B}^{-1}(\mu _{n,+}^{p})\prod_{j=k}^{p-1}\mathsf{B}(\mu
_{n,+}q^{j})\mathsf{A}^{-1}(\mu _{n,-})\mathsf{B}(\mu _{n,-})\mathsf{B}%
^{-1}(\mu _{n,+}q^{k})\mathsf{A}(\mu _{n,+}q^{k})\prod_{j=0}^{k-1}\mathsf{B}%
(\mu _{n,+}q^{j}),
\end{equation}%
Now by using the relations:%
\begin{equation}
\mathsf{A}^{-1}(\mu _{n,-})\mathsf{B}(\mu _{n,-})=\left( \mathsf{B}^{-1}(\mu
_{n,-})\mathsf{A}(\mu _{n,-})\right) ^{p-1},
\end{equation}%
we get the following representations: 
\begin{align}
\gamma _{k,n}& =\sum_{a=1}^{\mathsf{N}}\sum_{\substack{ \bar{\alpha}\equiv
\{\alpha _{1},...,\alpha _{\mathsf{N}}\}\in \mathbb{\mathsf{N}}_{0}^{\mathsf{%
N}}:  \\ \sum_{h=1}^{\mathsf{N}}\alpha _{h}=p-1}}\left[ 
\begin{array}{c}
p-1 \\ 
\bar{\alpha}%
\end{array}%
\right] \prod_{j=1}^{\mathsf{N}}\left( \prod_{h=0}^{\alpha _{j}-1}\frac{%
a(\eta _{j}q^{-h})}{(\mu _{n,-}q^{h}/\eta _{j}-\eta _{j}/\mu _{n,-}q^{h})}%
\right.  \notag \\
& \times \left. \prod_{i\neq j,i=1}^{\mathsf{N}}\prod_{h=\alpha _{i}-(\alpha
_{j}+\delta _{j,a})+1}^{\alpha _{i}}\frac{1}{(\eta _{j}q^{h}/\eta _{i}-\eta
_{i}/\eta _{j}q^{h})}\right) \frac{a(\eta _{a}q^{-\alpha _{a}})}{(\mu
_{n,+}q^{\alpha _{a}+r}/\eta _{a}-\eta _{a}/\mu _{n,+}q^{\alpha _{a}+r})} 
\notag \\
& \times \prod_{j=1}^{\mathsf{N}}\prod_{h=0}^{r-1}\frac{(\mu _{n,+}q^{\alpha
_{j}+h}/\eta _{j}-\eta _{j}/\mu _{n,+}q^{\alpha _{j}+h})}{(\mu
_{n,+}q^{h}/\eta _{j}-\eta _{j}/\mu _{n,+}q^{h})}\prod_{j=1}^{\mathsf{N}}%
\text{T}_{j}^{-(\alpha _{j}+\delta _{j,a})},
\end{align}%
from which our result follows.
\end{proof}

\section{Form factors of local operators\label{FF-loc-Op}}

In the following we will compute matrix elements (form factors) of the form\footnote{To simplify the notations in this introduction to Section \ref{FF-loc-Op} we
are omitting the index $k\in \{0,...,(p-1)/2\}$ in the transfer matrix
eigenstates which are required in the case of even chain.}:   
\begin{equation}
\langle t|O_{n}|t^{\prime }\rangle   \label{General-ME}
\end{equation}%
which by definition are the action of a covector  $\langle t|\in \mathcal{L}%
_{\mathsf{N}}$, a left $\mathsf{T}$-eigenstate defined in (\ref{eigenT-l}),
on the vector obtained by the action of the local operator $O_{n}$ on the
right $\mathsf{T}$-eigenstate $|t^{\prime }\rangle \in \mathcal{R}_{\mathsf{N%
}}$ defined in (\ref{eigenT-r}). Of course, these form factors depend on the
normalization of the states $\langle t|$ and $|t^{\prime }\rangle $ and then it
is worth pointing out that nevertheless we can use them to expand m-point
functions like:%
\begin{equation}
\frac{\langle t|O_{n_{1}}\cdots O_{n_{\text{m}}}|t\rangle }{\langle
t|t\rangle }.  \label{General-n-point-F}
\end{equation}%
Indeed, by definition these m-point functions are normalization independent
and by using m-1 times the decomposition of the identity $\left( \ref%
{Id-decomp}\right) $, we get the expansions:%
\begin{equation}
\frac{\langle t|O_{n_{1}}\cdots O_{n_{\text{m}}}|t\rangle }{\langle
t|t\rangle }=\sum_{t^{\left( 1\right) }(\lambda ),...,t^{\left( \text{m}%
-1\right) }(\lambda )\in \Sigma _{\mathsf{T}}}\frac{\langle
t|O_{n_{1}}|t^{(1)}\rangle \langle t^{(\text{m}-1)}|O_{n_{\text{m}%
}}|t\rangle \prod_{a=2}^{\text{m}-1}\langle t^{\left( a-1\right)
}|O_{n_{a}}|t^{\left( a\right) }\rangle }{\langle t|t\rangle \prod_{a=1}^{%
\text{m}-1}\langle t^{\left( a\right) }|t^{\left( a\right) }\rangle },
\label{FF-expansion}
\end{equation}%
where in the r.h.s there are exactly the  form factors $\left( \ref%
{General-ME}\right) $ that we are going to compute in this paper.

\subsection{Form factors of $\mathsf{u}_{n}$}

In this section we use the SOV-representations of local operators to give
some examples of completely resummed form factors.

\begin{proposition}
\label{FF-Prop1}Let $\langle t_{k}|$ and $|t_{k^{\prime }}^{\prime }\rangle $
be two eigenstates of the transfer matrix $\mathsf{T}(\lambda )$, then it
holds:%
\begin{equation}
\langle t_{k}|\mathsf{u}_{n}|t_{k^{\prime }}^{\prime }\rangle =\frac{\varphi
_{n}^{(t_{k})}}{\varphi _{n}^{(t_{k^{\prime }}^{\prime })}}\left( \delta
_{k,k^{\prime }+1}\right) ^{\mathtt{e}_{\mathsf{N}}}\det_{[\mathsf{N}]}(||%
\mathcal{U}_{a,b}^{\left( t,t^{\prime }\right) }(\mu _{n,+})||),  \label{u}
\end{equation}%
where $\varphi _{n}^{(t_{k})}$ and $\varphi _{n}^{(t_{k^{\prime }}^{\prime
})}$ are the eigenvalues of the shift operator $\mathsf{U}_{n}$ and $||%
\mathcal{U}_{a,b}^{\left( t,t^{\prime }\right) }(\lambda )||$ is the $[%
\mathsf{N}]\times \lbrack \mathsf{N}]$ matrix:%
\begin{align}
\mathcal{U}_{a,b}^{\left( t,t^{\prime }\right) }(\lambda )& \equiv \Phi
_{a,b+1/2}^{\left( t,t^{\prime }\right) }\text{ \ for \ }b\in \{1,...,[%
\mathsf{N}]-1\}, \\
\mathcal{U}_{a,[\mathsf{N}]}^{\left( t,t^{\prime }\right) }(\lambda )&
\equiv \frac{\left( \eta _{a}^{(0)}\right) ^{[\mathsf{N}]-1}}{\text{\textsc{k%
}}\eta _{\mathsf{N}}^{(0)}{}^{\mathtt{e}_{\mathsf{N}}}}\sum_{h=1}^{p}\frac{%
q^{([\mathsf{N}]-1)h}Q_{t^{\prime }}(\eta _{a}^{(h)})}{\omega _{a}(\eta
_{a}^{(h)})}\left[ \frac{\bar{Q}_{t}(\eta _{a}^{(h+1)})}{(\lambda /\eta
_{a}^{(h+1)}-\eta _{a}^{(h+1)}/\lambda )}\bar{a}(\eta _{a}^{(h)})\right. 
\notag \\
& +\left. \mathtt{e}_{\mathsf{N}}\bar{Q}_{t}(\eta _{a}^{(h)})\left( \frac{%
\lambda }{\prod_{j=1}^{\mathsf{N}}\xi _{j}}\left( \eta _{a}^{(h)}\right) ^{[%
\mathsf{N}]}q^{k^{\prime }}+\frac{\prod_{j=1}^{\mathsf{N}}\xi _{j}}{\lambda }%
\left( \eta _{a}^{(h)}\right) ^{-[\mathsf{N}]}q^{-k^{\prime }}\right) \right]
,
\end{align}%
where \textsc{k}$\equiv \prod_{n=1}^{\mathsf{N}}\kappa _{n}/i$.
\end{proposition}

\begin{proof}
The right SOV-representation of the operator $\mathsf{B}^{-1}(\lambda )%
\mathsf{A}(\lambda )$ reads:%
\begin{equation}
\mathsf{B}^{-1}(\lambda )\mathsf{A}(\lambda )|\mathbf{k}\rangle=\frac{\mathtt{e}_{\mathsf{N}}}{%
\eta _{\mathsf{N}}}\left( \frac{\lambda }{\eta _{\mathsf{A}}}\mathsf{T}_{%
\mathsf{N}}^{+}+\frac{\eta _{\mathsf{A}}}{\lambda }\mathsf{T}_{\mathsf{N}%
}^{-}\right)|\mathbf{k}\rangle +\sum_{a=1}^{[\mathsf{N}]}\mathsf{T}_{a}^{+}|\mathbf{k}\rangle\frac{\bar{a}(\eta
_{a})}{\text{\textsc{k}}\eta _{\mathsf{N}}^{\mathtt{e}_{\mathsf{N}}}(\lambda
/\eta _{a}q-\eta _{a}q/\lambda )}\prod_{b\neq a}\frac{1}{(\eta _{a}/\eta
_{b}-\eta _{b}/\eta _{a})},  \label{ExB-1A}
\end{equation}%
Let us denote with $[\mathsf{B}^{-1}(\lambda )\mathsf{A}(\lambda )]$ the
second term on the r.h.s. of (\ref{ExB-1A}). Then, we observe that from the
SOV-decomposition of the $\mathsf{T}$-eigenstates, we have:%
\begin{align}
\langle t_{k}|[\mathsf{B}^{-1}(\lambda )\mathsf{A}(\lambda )]|t_{k^{\prime
}}^{\prime }\rangle & =\left( \frac{\sum_{h_{\mathsf{N}}=1}^{p}q^{(k-1-k^{%
\prime })h_{\mathsf{N}}}}{p\eta^{(0)}_{\mathsf{N}}}\right) ^{\mathtt{e}_{%
\mathsf{N}}}\sum_{a=1}^{[\mathsf{N}]}\sum_{h_{1},...,h_{[\mathsf{N}%
]}=1}^{p}V(\left( \eta _{1}^{(h_{1})}\right) ^{2},...,\left( \eta _{\lbrack 
\mathsf{N}]}^{(h_{[\mathsf{N}]})}\right) ^{2})  \notag \\
& \times \prod_{b\neq a,b=1}^{[\mathsf{N}]}\frac{\eta
_{b}^{(h_{b})}Q_{t^{\prime }}(\eta _{b}^{(h_{b})})\bar{Q}_{t}(\eta
_{b}^{(h_{b})})}{\omega _{b}(\eta _{b}^{(h_{b})})((\eta
_{a}^{(h_{a})})^{2}-(\eta _{b}^{(h_{b})})^{2})}  \notag \\
& \times \frac{\bar{Q}_{t}(\eta _{a}^{(h_{a}+1)})Q_{t^{\prime }}(\eta
_{a}^{(h_{a})})}{\omega _{a}(\eta _{a}^{(h_{a})})\text{\textsc{k}}}\frac{%
\left( \eta _{a}^{(h_{a})}\right) ^{[\mathsf{N}]-1}\bar{a}(\eta
_{a}^{(h_{a})})}{(\lambda /\eta _{a}^{(h_{a}+1)}-\eta
_{a}^{(h_{a}+1)}/\lambda )},
\end{align}%
and so:%
\begin{align}
\langle t_{k}|[\mathsf{B}^{-1}(\lambda )\mathsf{A}(\lambda )]|t_{k^{\prime
}}^{\prime }\rangle & =\left( \frac{\delta _{k,k^{\prime }+1}}{\eta _{%
\mathsf{N}}^{(0)}}\right) ^{\mathtt{e}_{\mathsf{N}}}\sum_{a=1}^{[\mathsf{N}%
]}\sum_{\substack{ h_{1},...,h_{\mathsf{N}}=1  \\ {\small \overbrace{h_{a}%
\text{ is missing.}}}}}^{p}\underset{\ \ \ {\small \overbrace{(\text{We have
removed the row }a\text{.}})}}{\hat{V}_{a}(\left( \eta _{1}^{(h_{1})}\right)
^{2},...,\left( \eta _{\lbrack \mathsf{N}]}^{(h_{[\mathsf{N}]})}\right) ^{2})%
}  \notag \\
& \times \prod_{b\neq a,b=1}^{[\mathsf{N}]}\frac{\eta
_{b}^{(h_{b})}Q_{t^{\prime }}(\eta _{b}^{(h_{b})})\bar{Q}_{t}(\eta
_{b}^{(h_{b})})}{\omega _{b}(\eta _{b}^{(h_{b})})}  \notag \\
& \times (-1)^{[\mathsf{N}]+a}\sum_{h_{a}=1}^{p}\frac{\bar{Q}_{t}(\eta
_{a}^{(h_{a}+1)})Q_{t^{\prime }}(\eta _{a}^{(h_{a})})\left( \eta
_{a}^{(h_{a})}\right) ^{[\mathsf{N}]-1}\bar{a}(\eta _{a}^{(h_{a})})}{\omega
_{a}(\eta _{a}^{(h_{a})})\text{\textsc{k}}(\lambda /\eta
_{a}^{(h_{a}+1)}-\eta _{a}^{(h_{a}+1)}/\lambda )},
\end{align}%
bringing the sum over ($h_{1},...,\widehat{h_{a}},...,h_{[\mathsf{N}]}$)
inside the Vandermonde determinant $\hat{V}_{a}$, we have that the above
expression is just the expansion of the determinant:%
\begin{equation}
\langle t_{k}|[\mathsf{B}^{-1}(\lambda )\mathsf{A}(\lambda )]|t_{k^{\prime
}}^{\prime }\rangle =\left( \delta _{k,k^{\prime }+1}\right) ^{\mathtt{e}_{%
\mathsf{N}}}\det_{[\mathsf{N}]}(||\left[ \mathcal{U}_{a,b}^{\left(
t,t^{\prime }\right) }(\lambda )\right] ||),  \label{FF-Odd-part}
\end{equation}%
where $\left[ \mathcal{U}_{a,b}^{\left( t,t^{\prime }\right) }(\lambda )%
\right] $ coincides with $\Phi _{a,b+1/2}^{\left( t,t^{\prime }\right) }$
for \ $b\in \{1,...,[\mathsf{N}]-1\}$, while:%
\begin{equation}
\left[ \mathcal{U}_{a,[\mathsf{N}]}^{\left( t,t^{\prime }\right) }(\lambda )%
\right] \equiv \frac{\left( \eta _{a}^{(0)}\right) ^{[\mathsf{N}]-1}}{\text{%
\textsc{k}}\bar{\eta}_{\mathsf{N}}^{\mathtt{e}_{\mathsf{N}}}}\sum_{h=1}^{p}%
\frac{q^{([\mathsf{N}]-1)h}Q_{t^{\prime }}(\eta _{a}^{(h)})\bar{Q}_{t}(\eta
_{a}^{(h+1)})}{\omega _{a}(\eta _{a}^{(h)})(\lambda /\eta _{a}^{(h+1)}-\eta
_{a}^{(h+1)}/\lambda )}\bar{a}(\eta _{a}^{(h)}).
\end{equation}%
Now let us compute the matrix elements:%
\begin{align}
\langle t_{k}|\eta _{\mathsf{N}}^{-1}\eta _{\mathsf{A}}^{\mp }\mathsf{T}_{%
\mathsf{N}}^{\pm }|t_{k^{\prime }}^{\prime }\rangle & =\left( \frac{\sum_{h_{%
\mathsf{N}}=1}^{p}q^{(k-1-k^{\prime })h_{\mathsf{N}}}}{p\eta _{\mathsf{N}%
}^{(0)}\prod_{j=1}^{\mathsf{N}}\xi _{j}^{\pm 1}}\right) ^{\mathtt{e}_{%
\mathsf{N}}}\sum_{h_{1},...,h_{[\mathsf{N}]}=1}^{p}V(\left( \eta
_{1}^{(h_{1})}\right) ^{2},...,\left( \eta _{\lbrack \mathsf{N}]}^{(h_{[%
\mathsf{N}]})}\right) ^{2})  \notag \\
& \times \prod_{b=1}^{[\mathsf{N}]}\frac{\left( \eta _{b}^{(h_{b}+k^{\prime
})}\right) ^{\pm 1}Q_{t^{\prime }}(\eta _{b}^{(h_{b})})\bar{Q}_{t}(\eta
_{b}^{(h_{b})})}{\omega _{b}(\eta _{b}^{(h_{b})})},
\end{align}%
and so:%
\begin{equation}
\langle t_{k}|\eta _{\mathsf{N}}^{-1}\eta _{\mathsf{A}}^{\mp }\mathsf{T}_{%
\mathsf{N}}^{\pm }|t_{k^{\prime }}^{\prime }\rangle =\left( \frac{q^{\pm
k^{\prime }}\delta _{k,k^{\prime }+1}}{\eta _{\mathsf{N}}^{(0)}\prod_{j=1}^{%
\mathsf{N}}\xi _{j}^{\pm 1}}\right) ^{\mathtt{e}_{\mathsf{N}}}\det_{[\mathsf{%
N}]}(||\Phi _{a,b\pm 1/2}^{\left( t,t^{\prime }\right) }||).
\label{FF-Even-part}
\end{equation}%
By using the fact that $[\mathsf{N}]-1$ columns are common in the matrix of
formula (\ref{FF-Odd-part})\ and in those of (\ref{FF-Even-part}), we get
our result. Let us remark that the above result holds for any value of $%
\lambda $.
\end{proof}

\textbf{Remark 6.} It is worth pointing out that the form factors of $%
\mathsf{u}_{n}$ are written in terms of a determinant of a matrix whose
elements coincide with those of the scalar product, except for the last line
which is modified by the presence of the local operator. It is then
interesting to recall that a similar statement holds for the form factors of
the local operators in the XXZ spin 1/2 chain.

\subsection{Suitable operator basis for form factor computations}

In this section we introduce an operator basis which can be conveniently
used to describe all local operators. The interest toward this basis is due
to the fact that the form factors of its elements are simple being
represented by a determinant formula.

\subsubsection{Basis of elementary operators}

Let us introduce the following operators:%
\begin{equation}
\mathcal{O}_{a,k}\equiv \frac{\mathsf{B}(\eta _{a}^{(p+k-1)})\mathsf{B}(\eta
_{a}^{(p+k-2)})\cdots \mathsf{B}(\eta _{a}^{(k+1)})\mathsf{A}(\eta
_{a}^{(k)})}{p\eta _{\mathsf{N}}^{\mathtt{e}_{\mathsf{N}}(p-1)}\text{\textsc{%
k}}^{(p-1)}\prod_{b\neq a,b=1}^{[\mathsf{N}]}(Z_{a}/Z_{b}-Z_{b}/Z_{a})}\text{
\ with }k\in \{0,...,p-1\},
\end{equation}%
where the $\eta _{a}^{(k)}$ are fixed in Section \ref{SOV-Left}.

\begin{lemma}
The operators $\mathcal{O}_{a,k}$ satisfy the following properties:%
\begin{equation}
\mathcal{O}_{a,k}\mathcal{O}_{a,h}\text{ is non-zero if and only if }h=k-1,
\label{Prod-O-zeros}
\end{equation}%
and%
\begin{equation}
\mathcal{O}_{a,k}\mathcal{O}_{a,k-1}\cdots \mathcal{O}_{a,k+1-p}\mathcal{O}%
_{a,k-p}=\frac{\mathcal{A}(Z_{a})}{\prod_{b\neq a,b=1}^{[\mathsf{N}%
]}(Z_{a}/Z_{b}-Z_{b}/Z_{a})}\mathcal{O}_{a,k}.  \label{O-mean-value}
\end{equation}%
Moreover the following commutation relations hold:%
\begin{equation}
\eta _{\mathsf{A}}\mathcal{O}_{a,k}=q\mathcal{O}_{a,k}\eta _{\mathsf{A}},%
\text{ \ }[\eta _{\mathsf{N}},\mathcal{O}_{a,k}]=[\Theta ,\mathcal{O}%
_{a,k}]=0,
\end{equation}%
and 
\begin{equation}
\mathcal{O}_{a,k}\mathcal{O}_{b,h}=\frac{(\eta _{a}^{(k-h+1)}/\eta
_{b}^{(0)}-\eta _{b}^{(0)}/\eta _{a}^{(k-h+1)})}{(\eta _{a}^{(k-h-1)}/\eta
_{b}^{(0)}-\eta _{b}^{(0)}/\eta _{a}^{(k-h-1)})}\text{ }\mathcal{O}_{b,h}%
\mathcal{O}_{a,k}  \label{Com-O}
\end{equation}%
for $a\neq b\in \{1,...,[\mathsf{N}]\}$.
\end{lemma}

\begin{proof}
The first property follows from $\mathcal{B}(Z_{a})=0$, where $\mathcal{B}%
(\Lambda )$ is the average value of $\mathsf{B}(\lambda )$. By the
definition of $\mathcal{O}_{a,k}$ it is clear that:%
\begin{equation}
\langle \eta _{1}^{(k_{1})},...,\eta _{a}^{(h)},...,\eta _{\mathsf{N}}^{(k_{%
\mathsf{N}})}|\mathcal{O}_{a,k}=\frac{a(\eta _{a}^{(k)})\delta _{h,k}}{%
\prod_{b\neq a,b=1}^{\mathsf{N}}(\eta _{a}^{(k)}/\eta _{b}^{(k_{b})}-\eta
_{b}^{(k_{b})}/\eta _{a}^{(k)})}\langle \eta _{1}^{(k_{1})},...,\eta
_{a}^{(k-1)},...,\eta _{\mathsf{N}}^{(k_{\mathsf{N}})}|,  \label{O-Action}
\end{equation}%
so that the second property simply follows. To prove the last property we
have to use the Yang-Baxter commutation relation:%
\begin{equation}
(\lambda /\mu -\mu /\lambda )\mathsf{A}(\lambda )\mathsf{B}(\mu )=(\lambda
/q\mu -\mu q/\lambda )\mathsf{B}(\mu )\mathsf{A}(\lambda )+(q-q^{-1})\mathsf{%
B}(\lambda )\mathsf{A}(\mu )  \label{AB-Yang-Baxter}
\end{equation}%
we have before to move the $\mathsf{A}(\eta _{a}^{(k)})$ to the right
through all the $\mathsf{B}(\eta _{a}^{(j)})$, remarking that only the first
term of the r.h.s of (\ref{AB-Yang-Baxter}) survives, and after to move the $%
\mathsf{A}(\eta _{a}^{(h)})$ to the left.
\end{proof}

Let us introduce now the following monomials which we will call \textit{%
elementary operators:}%
\begin{equation}
\mathcal{E}_{(k,k_{0})^{\mathtt{e}_{\mathsf{N}%
}},(a_{1},k_{1}),...,(a_{r},k_{r})}^{(\alpha _{1},...,\alpha _{r})}\equiv
\eta _{\mathsf{N}}^{-\mathtt{e}_{\mathsf{N}}k}\left( \frac{\Theta }{\eta _{%
\mathsf{A}}}\right) ^{\mathtt{e}_{\mathsf{N}}k_{0}}\mathcal{O}%
_{a_{1},k_{1}}^{(\alpha _{1})}\cdots \mathcal{O}_{a_{r},k_{r}}^{(\alpha
_{r})},  \label{O-basis}
\end{equation}%
where $\sum_{h=1}^{r}\alpha _{h}\leq p,$ $k,k_{i}\in \{0,...,p-1\},\ \
a_{i}<a_{j}\in \{1,...,[\mathsf{N}]\}$ \ for $i<j\in \{1,...,[\mathsf{N}]\}$
and:%
\begin{equation}
\mathcal{O}_{a,k}^{(\alpha )}\equiv \mathcal{O}_{a,k}\mathcal{O}%
_{a,k-1}\cdots \mathcal{O}_{a,k+1-\alpha },\text{ with }\alpha \in
\{1,...,p\}.
\end{equation}%
Then the following lemma holds:

\begin{lemma}
For any $n\in \{1,...,\mathsf{N}\}$, the set of the elementary operators
dressed by the shift operator $\mathsf{U}_{n}$:%
\begin{equation}
\mathsf{U}_{n}\mathcal{E}_{(k,k_{0})^{\mathtt{e}_{\mathsf{N}%
}},(a_{1},k_{1}),...,(a_{r},k_{r})}^{(\alpha _{1},...,\alpha _{r})}\mathsf{U}%
_{n}^{-1},
\end{equation}%
is a basis in the space of the local operators at the quantum site $n$.
\end{lemma}

\begin{proof}
The space of the local operators in site $n$ is generated by $\mathsf{u}%
_{n}^{k}$ and $\mathsf{v}_{n}^{k}$ for $k\in \{1,...,p-1\}$. Proceeding as
done in Proposition \ref{IPS} we have in particular the possibility to show
that an alternative local basis is defined by the operators: 
\begin{align}
\mathsf{u}_{n}^{k}& =\mathsf{U}_{n}\left( \mathsf{B}^{-1}(\mu _{n,+})\mathsf{%
A}(\mu _{n,+})\right) ^{k}\mathsf{U}_{n}^{-1},  \label{U-B-basis1} \\
\tilde{\beta}_{k,n}& =\mathsf{U}_{n}\left( \mathsf{B}^{-1}(\mu _{n,+})%
\mathsf{A}(\mu _{n,+})\right) ^{k}\mathsf{B}^{-1}(\mu _{n,-})\mathsf{A}(\mu
_{n,-})\left( \mathsf{B}^{-1}(\mu _{n,+})\mathsf{A}(\mu _{n,+})\right)
^{p-1-k}\mathsf{U}_{n}^{-1}  \label{U-B-basis2}
\end{align}%
for $k\in \{1,...,p-1\}$. Then to prove the lemma we just have to show that
the above operators are linear combinations of those defined in (\ref%
{O-basis}). Note that for $\lambda ^{p}\neq Z_{a}$ with $a\in \{1,...,[%
\mathsf{N}]\}$, the operator $\mathsf{B}^{-1}(\lambda )$ is invertible and
by the centrality of the average values we can write: 
\begin{equation}
\mathsf{B}^{-1}(\lambda )\mathsf{A}(\lambda )=\frac{\mathsf{B}(\lambda
q^{p-1})\mathsf{B}(\lambda q^{p-2})\cdots \mathsf{B}(\lambda q)\mathsf{A}%
(\lambda )}{\mathcal{B}(\Lambda )}.
\end{equation}%
Now the operator $\mathsf{B}(\lambda q^{p-1})\mathsf{B}(\lambda
q^{p-2})\cdots \mathsf{B}(\lambda q)\mathsf{A}(\lambda )$ is an even Laurent
polynomial of degree:%
\begin{equation*}
(p-1)[\mathsf{N}]+\mathsf{N}-1+\mathtt{e}_{\mathsf{N}}=\left\{ 
\begin{array}{l}
p\mathsf{N}-1\text{ \ \ \ \ \ \ \ \ \ for }\mathsf{N}\text{ odd} \\ 
p(\mathsf{N}-1)+1\text{ for }\mathsf{N}\text{ even}%
\end{array}%
\right.
\end{equation*}%
in $\lambda $. So for $\mathsf{N}$ odd to completely characterize it we have
to fix its value in $p\mathsf{N}$ distinguished points and we are free to
chose these points coinciding with the zeros of the operator $\mathsf{B}%
(\lambda )$. For $\mathsf{N}$ even, we have to add to the $\mathsf{B}$-zeros
the values at the infinity, so that by using the corresponding interpolation
formula for $\mathsf{B}(\lambda q^{p-1})\mathsf{B}(\lambda q^{p-2})\cdots 
\mathsf{B}(\lambda q)\mathsf{A}(\lambda )$, we derive:%
\begin{equation}
\mathsf{B}^{-1}(\lambda )\mathsf{A}(\lambda )=\frac{\mathtt{e}_{\mathsf{N}}}{%
\eta _{\mathsf{N}}}\left( \frac{\lambda \Theta }{\eta _{\mathsf{A}}}+\frac{%
\eta _{\mathsf{A}}}{\lambda \Theta }\right) +\frac{1}{\eta _{\mathsf{N}}^{%
\mathtt{e}_{\mathsf{N}}}}\sum_{a=1}^{[\mathsf{N}]}\sum_{k=0}^{p-1}\frac{%
\mathcal{O}_{a,k}}{(\lambda /\eta _{a}^{(k)}-\eta _{a}^{(k)}/\lambda )}.
\end{equation}%
From the previous formula and the representation (\ref{U-B-basis1})-(\ref%
{U-B-basis2}), we have that the local operators $\mathsf{u}_{n}^{k}$ and $%
\tilde{\beta}_{k,n}$ are linear combinations of the monomials $\mathsf{U}%
_{n}\eta _{\mathsf{N}}^{-\mathtt{e}_{\mathsf{N}}h}\left( \frac{\Theta }{\eta
_{\mathsf{A}}}\right) ^{\mathtt{e}_{\mathsf{N}}h_{0}}\mathcal{O}%
_{a_{1},h_{1}}\cdots \mathcal{O}_{a_{s},h_{s}}\mathsf{U}_{n}^{-1}$ for $%
s\leq p$, $a_{i}\in \{1,...,[\mathsf{N}]\}$ and $h,h_{i}\in \{0,...,p-1\}$.
For any monomial $\mathcal{O}_{a_{1},h_{1}}\cdots \mathcal{O}_{a_{s},h_{s}}$
we can use the commutation rules (\ref{Com-O}) to rewrite it in a way that
operators with the same index $a$ are adjacent, we can order them in a way
that $a_{i}<a_{j}$\ for $i<j\in \{1,...,[\mathsf{N}]\}$\ and we can apply
the rule (\ref{Prod-O-zeros}) to say if the monomial is zero or not.
Finally, by using the property (\ref{O-mean-value}), we have: 
\begin{equation}
\mathcal{O}_{a,k}^{(p+\alpha )}=\frac{\mathcal{A}(Z_{a})}{\prod_{b\neq
a,b=1}^{\mathsf{N}}(Z_{a}/Z_{b}-Z_{b}/Z_{a})}\mathcal{O}_{a,k}^{(\alpha )},
\end{equation}%
and so it is clear that all the non-zero monomials $\mathcal{O}%
_{a_{1},h_{1}}\cdots \mathcal{O}_{a_{s},h_{s}}$ can be written in the form (%
\ref{O-basis}).
\end{proof}

\subsubsection{Form factors of elementary operators}

As anticipated the interest in the above definition of elementary operators
is the simplicity of their form factors:

\begin{lemma}
Let $\langle t_{k}|$ and $|t_{k^{\prime }}^{\prime }\rangle $ be two
eigenstates of the transfer matrix $\mathsf{T}(\lambda )$, then it holds:%
\begin{equation}
\langle t_{k}|\mathcal{E}_{(h,h_{0})^{\mathtt{e}_{\mathsf{N}%
}},(a_{1},h_{1}),...,(a_{r},h_{r})}^{(\alpha _{1},...,\alpha
_{r})}|t_{k^{\prime }}^{\prime }\rangle =\frac{\delta _{k,k^{\prime }+h}^{%
\mathtt{e}_{\mathsf{N}}}q^{\mathtt{e}_{\mathsf{N}}h_{0}k^{\prime }}}{\eta _{%
\mathsf{N}}^{(0)h\mathtt{e}_{\mathsf{N}}}\prod_{j=1}^{\mathsf{N}}\xi _{j}^{%
\mathtt{e}_{\mathsf{N}}h_{0}}}\,\mathsf{f}_{(\mathtt{e}_{\mathsf{N}%
}h_{0},\{\alpha \},\{a\})}\det_{[\mathsf{N}]+rp-g}(||\text{\textsc{O}}%
_{a,b}^{(\mathtt{e}_{\mathsf{N}}h_{0},\{\alpha \},\{a\})}||),
\end{equation}%
where $||$\textsc{O}$_{a,b}^{(\mathtt{e}_{\mathsf{N}}h_{0},\{\alpha
\},\{a\})}||$ is the $\left( [\mathsf{N}]+rp-g\right) \times \left( \lbrack 
\mathsf{N}]+rp-g\right) $ matrix of elements:%
\begin{align}
\text{\textsc{O}}_{a,\sum_{h=1}^{m-1}(p-\alpha _{h}+1)+j_{m}}^{(\mathtt{e}_{%
\mathsf{N}}h_{0},\{\alpha \},\{a\})}& \equiv \left( \eta
_{a_{m}}^{(h_{m}+j_{m})}\right) ^{4(a-1)}\text{ \ for }j_{m}\in
\{0,...,p-\alpha _{m}\},\text{ \ }m\in \{1,...,r\},\text{ } \\
\text{\textsc{O}}_{a,\sum_{h=1}^{r}(p-\alpha _{h}+1)+i}^{(\mathtt{e}_{%
\mathsf{N}}h_{0},\{\alpha \},\{a\})}& \equiv \Phi _{b_{i},a+(\mathtt{e}_{%
\mathsf{N}}h_{0}+g)/2}^{\left( t,t^{\prime }\right) }\text{ \ \ \ \ \ \ \ \
\ \ for }i\in \{1,...,[\mathsf{N}]-r\},\text{ \ \ \ }g\equiv
\sum_{h=1}^{r}\alpha _{h},
\end{align}%
for any $a\in \{1,...,[\mathsf{N}]+rp-g\}$. Here, we have defined $%
\{b_{1},...,b_{[\mathsf{N}]-r}\}\equiv \{1,...,\mathsf{[\mathsf{N}]}%
\}\backslash \{a_{1},...,a_{r}\}$ with elements ordered by $b_{i}<b_{j}$ for 
$i<j$ and%
\begin{align}
\mathsf{f}_{(\mathtt{e}_{\mathsf{N}}h_{0},\{\alpha \},\{a\})}& \equiv \frac{%
\prod_{i=1}^{r}Q_{t^{\prime }}(\eta _{a_{i}}^{(h_{i}-\alpha _{i})})\bar{Q}%
_{t}(\eta _{a_{i}}^{(h_{i})})\frac{\left( \eta _{a_{i}}^{(h_{i})}\right) ^{%
\mathtt{e}_{\mathsf{N}}h_{0}+\alpha _{i}\left( [\mathsf{N}]-r\right) }}{%
\omega _{a_{i}}(\eta _{a_{i}}^{(h_{i})})}\prod_{h=0}^{\alpha _{i}-1}a(\eta
_{a_{i}}^{(h_{i}-h)})}{\prod_{i=1}^{r}\prod_{h=0}^{\alpha
_{i}-1}\prod_{j=1}^{i-1}(\frac{\eta _{a_{i}}^{(h_{i}+\alpha _{i}-h)}}{\eta
_{a_{j}}^{(h_{j})}}-\frac{\eta _{a_{j}}^{(h_{j})}}{\eta
_{a_{i}}^{(h_{i}+\alpha _{i}-h)}})\prod_{j=i+1}^{r}(\frac{\eta
_{a_{i}}^{(h_{i})}}{\eta _{a_{j}}^{(h_{j}+h)}}-\frac{\eta
_{a_{j}}^{(h_{j}+h)}}{\eta _{a_{i}}^{(h_{i})}})}  \notag \\
& \times \frac{(-1)^{\sum_{i=1}^{r}(a_{i}-i)}\prod_{i=1}^{r}q^{-\left( [%
\mathsf{N}]-r\right) \alpha _{i}(\alpha _{i}-1)/2}V((\eta
_{a_{1}}^{(h_{1})})^{2},...,(\eta _{a_{r}}^{(h_{r})})^{2})}{%
\prod_{i=1}^{r}\prod_{j=1}^{[\mathsf{N}]-r}(Z_{a_{i}}^{2}-Z_{b_{j}}^{2})V((%
\eta _{a_{1}}^{(h_{1})})^{2},...,(\eta _{a_{1}}^{(h_{1}+p-\alpha
_{1})})^{2},...,(\eta _{a_{r}}^{(h_{r})})^{2}...,(\eta
_{a_{r}}^{(h_{r}+p-\alpha _{r})})^{2}))},
\end{align}%
where $V(x_{1},...,x_{\mathsf{N}})\equiv \prod_{1\leq b<a\leq \mathsf{N}%
}(x_{a}-x_{b})$ is the Vandermonde determinant.
\end{lemma}

\begin{proof}
The operator $\eta _{\mathsf{N}}^{-\mathtt{e}_{\mathsf{N}}h}\left( \frac{%
\Theta }{\eta _{\mathsf{A}}}\right) ^{\mathtt{e}_{\mathsf{N}}h_{0}}$ act in
the following way on the state $\langle t_{k}|$:%
\begin{align}
\langle t_{k}|\eta _{\mathsf{N}}^{-\mathtt{e}_{\mathsf{N}}h}\left( \frac{%
\Theta }{\eta _{\mathsf{A}}}\right) ^{\mathtt{e}_{\mathsf{N}}h_{0}}& = 
\notag \\
& =\left( \frac{q^{h_{0}(k-h)}}{(\eta _{\mathsf{N}}^{(0)})^{2}\prod_{j=1}^{%
\mathsf{N}}\xi _{j}^{h_{0}}}\right) ^{\mathtt{e}_{\mathsf{N}%
}}\sum_{k_{1},...,k_{\mathsf{N}}=1}^{p}\left( \frac{q^{(k-h)h_{\mathsf{N}}}}{%
p^{1/2}}\right) ^{\mathtt{e}_{\mathsf{N}}}\prod_{a=1}^{[\mathsf{N}]}(\eta
_{a}^{(k_{a})})^{h_{0}\mathtt{e}_{\mathsf{N}}}\bar{Q}_{t}(\eta
_{a}^{(k_{a})})  \notag \\
& \times \prod_{1\leq b<a\leq \lbrack \mathsf{N}]}((\eta
_{a}^{(k_{a})})^{2}-(\eta _{b}^{(k_{b})})^{2})\frac{\langle \eta
_{1}^{(k_{1})},...,\eta _{\mathsf{N}}^{(k_{\mathsf{N}})}|}{\prod_{b=1}^{[%
\mathsf{N}]}\omega _{b}(\eta _{b}^{(k_{b})})}.
\end{align}%
From the formula (\ref{O-Action}), it follows:%
\begin{equation}
\langle \eta _{1}^{(k_{1})},...,,...,\eta _{a_{i}}^{(f)},...,\eta _{\mathsf{N%
}}^{(k_{\mathsf{N}})}|\mathcal{O}_{a_{i},h_{i}}^{(\alpha _{i})}=\frac{%
\prod_{h=0}^{\alpha _{i}-1}a(\eta _{a_{i}}^{(h_{i}-h)})\delta
_{f,h_{i}}\langle \eta _{1}^{(k_{1})},...,,...,\eta _{a_{i}}^{(h_{i}-\alpha
_{i})},...,\eta _{\mathsf{N}}^{(k_{\mathsf{N}})}|}{\prod_{b\neq a_{i},b=1}^{[%
\mathsf{N}]}\prod_{h=0}^{\alpha _{i}-1}(\eta _{a_{i}}^{(h_{i}-h)}/\eta
_{b}^{(k_{b})}-\eta _{b}^{(k_{b})}/\eta _{a_{i}}^{(h_{i}-h)})}.
\label{O-a-Action}
\end{equation}
So we can compute also the action of $\mathcal{O}_{a_{1},h_{1}}^{(\alpha
_{1})}\cdots \mathcal{O}_{a_{r},h_{r}}^{(\alpha _{r})}$ just taking into
account the order of the operators in the monomial which leads by the scalar
product formula to:%
\begin{align}
\langle t_{k}|\mathcal{E}_{(h,h_{0})^{\mathtt{e}_{\mathsf{N}%
}},(a_{1},h_{1}),...,(a_{r},h_{r})}^{(\alpha _{1},...,\alpha
_{r})}|t_{k^{\prime }}^{\prime }\rangle & =\left( \frac{q^{h_{0}(k-h)}}{%
(\eta _{\mathsf{N}}^{(0)})^{h}\prod_{j=1}^{\mathsf{N}}\xi _{j}^{h_{0}}}%
\right) ^{\mathtt{e}_{\mathsf{N}}}\sum_{k_{1},...,k_{\mathsf{N}%
}=1}^{p}\left( \frac{q^{\left[ (k-h)-k^{\prime }\right] k_{\mathsf{N}}}}{p}%
\right) ^{\mathtt{e}_{\mathsf{N}}}\prod_{a=1}^{[\mathsf{N}]}(\eta
_{a}^{(k_{a})})^{h_{0}\mathtt{e}_{\mathsf{N}}}  \notag \\
& \times \prod_{i=1}^{r}\frac{\prod_{h=0}^{\alpha _{i}-1}a(\eta
_{a_{i}}^{(h_{i}-h)})\delta _{k_{a_{i}},h_{i}}}{\prod_{j=1}^{[\mathsf{N}%
]-r}\prod_{h=0}^{\alpha _{i}-1}(\eta _{a_{i}}^{(h_{i}-h)}/\eta
_{b_{j}}^{(k_{b_{j}})}-\eta _{b_{j}}^{(k_{b_{j}})}/\eta _{a_{i}}^{(h_{i}-h)})%
}  \notag \\
& \times \prod_{i=1}^{r}\prod_{h=0}^{\alpha _{i}-1}\frac{\prod_{j=i+1}^{r}(%
\eta _{a_{i}}^{(h_{i}-h)}/\eta _{a_{j}}^{(h_{j})}-\eta
_{a_{j}}^{(h_{j})}/\eta _{a_{i}}^{(h_{i}-h)})^{-1}}{\prod_{j=1}^{i-1}(\eta
_{a_{i}}^{(h_{i}-h)}/\eta _{a_{j}}^{(h_{j}-\alpha _{j})}-\eta
_{a_{j}}^{(h_{j}-\alpha _{j})}/\eta _{a_{i}}^{(h_{i}-h)})}  \notag \\
& \times \prod_{j=1}^{[\mathsf{N}]-r}\frac{Q_{t^{\prime }}(\eta
_{b_{j}}^{(k_{b_{j}})})\bar{Q}_{t}(\eta _{b_{j}}^{(k_{b_{j}})})}{\omega
_{b_{j}}(\eta _{b_{j}}^{(k_{b_{j}})})}\prod_{i=1}^{r}\frac{Q_{t^{\prime
}}(\eta _{a_{i}}^{(h_{i}-\alpha _{i})})\bar{Q}_{t}(\eta _{a_{i}}^{(h_{i})})}{%
\omega _{a_{i}}(\eta _{a_{i}}^{(h_{i})})}  \notag \\
&  \times V((\eta
_{1}^{(h_{1})})^{2},...,(\eta _{\lbrack \mathsf{N}]}^{(h_{[\mathsf{N}%
]})})^{2}).
\end{align}%
Let us remark that the sum $\sum_{k_{1},...,k_{\mathsf{N}}=1}^{p}$ reduces
to $\delta _{k,k^{\prime }+h}^{\mathtt{e}_{\mathsf{N}}}$ times the sum $%
\sum_{k_{b_{1}},...,k_{b_{\mathsf{[N]}-r}}=1}^{p}$ for the presence of the $%
\prod_{i=1}^{r}\delta _{k_{a_{i}},h_{i}}$. Now we multiply each term of the
sum by: 
\begin{align}
1& =\prod_{\epsilon =\pm 1}\prod_{i=1}^{r}\prod_{j=1}^{[\mathsf{N}%
]-r}\prod_{h=-p+\alpha _{i}}^{-1}((\eta {}_{a_{i}}^{(h_{i}-h)})^{2}-(\eta
_{b_{j}}^{(k_{b_{j}})})^{2})^{\epsilon }  \notag \\
& \times \left( \frac{V((\eta _{a_{1}}^{(h_{1})})^{2},...,(\eta
_{a_{1}}^{(h_{1}+p-\alpha _{1})})^{2},...,(\eta
_{a_{r}}^{(h_{r})})^{2}...,(\eta _{a_{r}}^{(h_{r}+p-\alpha _{r})})^{2}))}{%
V((\eta _{a_{1}}^{(h_{1})})^{2},...,(\eta _{a_{r}}^{(h_{r})})^{2})}\right)
^{\epsilon }
\end{align}%
here the power $+1$ leads to the construction of the Vandermonde determinant:%
\begin{equation}
V(\underset{p-\alpha _{1}+1\text{ columns}}{\underbrace{(\eta
_{a_{1}}^{(h_{1})})^{2},...,(\eta _{a_{1}}^{(h_{1}+p-\alpha _{1})})^{2}}}%
,...,\underset{p-\alpha _{r}+1\text{ columns}}{\underbrace{(\eta
_{a_{r}}^{(h_{r})})^{2},...,(\eta _{a_{r}}^{(h_{r}+p-\alpha _{r})})^{2}}},%
\underset{[\mathsf{N}]-r\text{ columns}}{\underbrace{(\eta
_{b_{1}}^{(k_{b_{1}})})^{2},...,(\eta _{b_{[\mathsf{N}]-r}}^{(k_{b_{[\mathsf{%
N}]-r}})})^{2}}}),
\end{equation}%
and the sum $\sum_{k_{b_{1}},...,k_{b_{\mathsf{[N]}-r}}=1}^{p}$ becomes sum
over columns and after some algebra we get our formula.
\end{proof}

It is interesting to point out that the last $[\mathsf{N}]-r$ columns of the
matrix $||\text{\textsc{O}}_{a,b}^{(\mathtt{e}_{\mathsf{N}}h_0,\{\alpha
\},\{a\})}||$ are just those of the scalar product.

\textbf{Remark 7. } For the similarity of the model and representations
considered, it is natural to cite the series of works \cite%
{GIPST07,GIPS06,GIPST08,GIPS09}. There, in the framework of cyclic
SOV-representations, first results on the matrix elements of local operators
appear. However, it is worth saying that these quantities\ are there
computed only for the restriction\footnote{%
It can be compared to the restriction to the case $q^{2}=1$ for the even
sine-Gordon chain.} of the $\tau _{2}$-model\ to the generalized Ising
model. In particular, the matrix elements of $\mathsf{u}_{1}$ are computed
and the results are not presented in a determinant form.

\section{Conclusion and outlook}

\subsection{Results}

In this article we have considered the lattice sine-Gordon model in 
cyclic representations and we have solved in this case two fundamental
problems for the computation of matrix elements of local operators:

\begin{itemize}
\item Scalar products: determinant of $\mathsf{N}\times\mathsf{N}$
matrices whose matrix elements are sums over the spectrum of each quantum
separate variable of the product of the coefficients of states, this being
for all the left/right separate states in the SOV-basis.

\item Inverse problem solution: reconstruction of all local operators in
terms of standard Sklyanin's quantum separate variables.
\end{itemize}

Further, we have shown how these results lead to the computation of matrix
elements of all local operators. At first, standard\footnote{%
Coinciding with the operator-zeros of one of the Yang-Baxter algebra
generators, like $B(\lambda )$ or $C(\lambda )$.} Sklyanin's quantum
separate variables are suitable for solving the transfer matrix spectral
problem. Indeed, the transfer matrix spectrum (eigenvalues \& eigenstates)
admits a simple and complete characterization in terms of Baxter-equation
solutions in this SOV-basis. Then the inverse problem solution allows to
write the action of any local operator on transfer matrix eigenstates as
finite sums of separate states in the SOV-basis. Hence, the matrix elements
of any local operator are written as finite sums of determinants of the
resulting scalar product formulae.

We have explicitly developed this program characterizing the matrix elements
of the local operators $\mathsf{u}_{n}$ and $\alpha _{n}$ by one determinant
formulae in terms of matrices obtained by modifying a single row in the
scalar product matrices. Moreover, we have constructed an operator basis
whose matrix elements are in turn written by one determinant formulae. The
matrices involved have rows which coincide with those of the scalar product
matrix or with those of the Vandermonde matrix computed in the spectrum of
the separate variables.

\subsection{Comparison with previous SOV-results}

In the literature of quantum integrable models there exist several results
on matrix elements of local operators which can be traced back to
applications of separation of variable methods. In this section, we try to
recall the most relevant ones as they allow for an explicit comparison with
our results. It leads to a universal picture emerging in the
characterization of matrix elements by SOV-methods.

\subsubsection{On the reconstruction of local operators}

One important motivation for our work was to introduce a well defined setup
which allows to solve the longstanding problem of the identification of
local operators in the continuum sine-Gordon model thanks to the
reconstructions achieved on the lattice. Then, it should also allow for the
identification of form factor solutions of the continuous theory by
implementing well defined limits from our lattice formulae.

Even if methodologically different, it is worth recalling the semi-classical
reconstruction presented by Babelon, Bernard and Smirnov in \cite{BabBS96}
for chiral local operators of the restricted sine-Gordon model (in the
infinite volume) at the reflectionless points, $\beta ^{2}=1/(1+\nu )$ with $%
\nu \in \mathbb{Z}^{\geq 0}$. The classical sine-Gordon model admits a SOV
description: each $n$-soliton solution $\varphi (x,t)$\ of the equation of
motion can be represented in terms of $n$-separate variables A$_{j}$, which
in the BBS choice \cite{BabBS96} lead to the representations:%
\begin{equation}
e^{i\varphi }=\prod_{j=1}^{n}\frac{\text{A}_{j}}{\text{B}_{j}},
\label{SemiCl-Recontruction}
\end{equation}%
where the B$_{j}$ are integrals of motion. The formula (\ref%
{SemiCl-Recontruction}) represents classically a SOV reconstruction of the
local fields when restricted to the $n$-soliton sector. In \cite{BabBS96},
this reconstruction has been extended to the quantum model in each $n$%
-solitons sector by quantizing the separate variables\footnote{%
A fundamental point is the introduction of appropriate Hermitian conjugation
properties and the characterization of the spectrum of the Weyl algebra
generators.} A$_{j}$ and the conjugate momenta as operators which generate $%
n $ independent Weyl algebras with parameter $\tilde{q}=e^{i\pi \frac{\beta
^{2}}{1-\beta ^{2}}}=e^{i\frac{\pi }{v}}$. \ This extension and the
consequent identifications of primary fields and their chiral descendants in
the perturbed minimal models M$_{1,1+\nu }$ are justified by the following
indirect but strong arguments: a) The $n$-multiple integrals of the form
(36) in \cite{BabBS96} which represent the $n$-solitons to $n$-solitons form
factors\footnote{%
Note that these are solutions of the form factor equations and so they
surely represent local fields in the S-matrix formulation of the restricted
sine-Gordon model.} of chiral left operators at the reflectionless points
are reproduced from the semi-classical limit. b) The counting of these form
factor solutions allows the reconstruction of the chiral characters of M$%
_{1,1+\nu }$\ \cite{BabBS97}. Further support to (\ref{SemiCl-Recontruction}%
) was given by Smirnov's work on semi-classical form factors\footnote{%
On the basis of this last Smirnov's work, Lukyanov has
introduced his conjecture for the finite temperature expectations values of
exponential fields in finite volume for the shG-model \cite{Luk01}.} of the
continuous KdV model in finite volume \cite{Smi98}; there the form factors
of \cite{BabBS96} were reproduced by taking the infinite volume limit of the
KdV semi-classical ones.

Let us remark that in our lattice regularization of the sine-Gordon model,
choosing as quantum separate variables standard\footnote{%
The same is true if we take the products of the operator zeros of 
\textsf{C}$(\lambda )$ but also of \textsf{A}$(\lambda )$ and \textsf{D}$%
(\lambda )$, i.e. for all possibilities to construct the SOV representations
by the simplest Sklyanin's method.} Sklyanin's ones, the reconstruction of the
exponential fields is not of the simple form given in (\ref%
{SemiCl-Recontruction}). Then, the following question is relevant: is it
possible to find a SOV representation of the quantum lattice sine-Gordon
model where the exponential fields are simply written as a product of
generators of the SOV representations?

A natural idea can be to implement a change of basis in the quantum separate
variables from Sklyanin's ones to a new set; an interesting example of this
approach was used by Babelon in \cite{Bab04} which has provided a simple
reconstruction of the lattice quantum Toda local operators in terms of a set
of quantum separate variables defined by a change of variables from the 
Sklyanin's ones. However, it is worth pointing out that, for a new SOV
representation to be really useful for the computation of matrix elements,
it should not only give a simple reconstruction of the local operators but
also keep the solution of the transfer matrix spectral problem and the
scalar product formulae as simple as for original Sklyanin's variables. Let
us comment that a reconstruction like (\ref{SemiCl-Recontruction}) can be
formally derived at the quantum level implementing the special limit 
\begin{equation}
\kappa_{n}/i\rightarrow +\infty  \label{L1}
\end{equation}%
on the following reconstruction formula of the lattice sine-Gordon model:%
\begin{equation}
\frac{(q^{-1}\mathsf{v}_{n}^{2}+\kappa _{n}^{2})}{\left( q^{-1}\mathsf{v}%
_{n}^{2}\kappa _{n}^{2}+1\right) }=\mathsf{U}_{n}\mathsf{A}^{-1}(\mu _{n,-})%
\mathsf{B}(\mu _{n,-})\mathsf{B}^{-1}(\mu _{n,+})\mathsf{A}(\mu _{n,+})%
\mathsf{U}_{n}^{-1}\ .  \label{Oota^2}
\end{equation}%
The result for an even chain reads 
\begin{equation}
\mathsf{v}_{n}^{-2h}=\frac{\Theta ^{2h}}{\prod_{a\neq n,a=1}^{\mathsf{N%
}}\xi _{a}^{2h}}\mathsf{U}_{n}\prod_{a=1}^{\mathsf{N}-1}\eta _{a}^{2h}%
\mathsf{U}_{n}^{-1},\text{ \ }h\in \{1,..,p-1\}  \label{Formula+}
\end{equation}%
in terms of Sklyanin's separate variables. It is possible to argue that the
previous limit can be consistently interpreted as a chiral deformation of
the lattice sine-Gordon model to chiral KdV models \cite{NovMPZ84,FadV94}.
In a future publication, we will analyze the cyclic representations\footnote{%
Here, we are referring to compact representations of chiral KdV models where
the generators of the local Weyl algebras are unitary operators. The
spectrum of the non-compact versions was instead analyzed by SOV and
Q-operator method in \cite{BytT09}.} of these chiral KdV models showing our
statement on the reconstruction formulae and computing the matrix elements.
It is worth pointing out that the form factors of the r.h.s. of (\ref%
{Formula+}) are trivial to compute in our SOV framework and are written by
one determinant formulae which differ w.r.t. the scalar product only for $%
h\in \{1,..,p-1\}$ rows.

\subsubsection{On the matrix elements of local operators}

In the case of the quantum integrable Toda chain \cite{Skl85}, Smirnov \cite%
{Smi98} has derived in the framework of Sklyanin's SOV determinant formulae
for the matrix elements of a conjectured basis of local operators which look
very similar to our formulae. The main difference is due to the different
nature of the spectrum of the quantum separate variables in the two models.
In fact, in the case of the lattice Toda model, Sklyanin's measure is
continuous (continuous SOV-spectrum) while it is discrete in the case of the
cyclic lattice sine-Gordon model. The elements of the matrices whose
determinants give the form factor formulae are then expressed as
\textquotedblleft convolutions\textquotedblright , over the spectrum of the
separate variables, of Baxter equations solutions plus contributions coming
from the local operators. In the case of Smirnov's formulae they are true
integrals, the SOV-spectrum being continuous, while in our formulae they are
\textquotedblleft discrete convolutions\textquotedblright\ , the
SOV-spectrum being discrete. Let us comment that the need to conjecture%
\footnote{%
The consistency of this conjecture is there verified by a counting argument
based on the existence of an appropriate set of null conditions for the
\textquotedblleft integral convolutions\textquotedblright .} the form of a
basis of local operators in \cite{Smi98} is due to the lack of a direct
reconstruction of local operators in terms of Sklyanin's separate variables.

In the case of the infinite volume quantum sine-Gordon field theory, the
form factors of local operators \cite{Smi92b} have also a form similar to
the one predicted by SOV.\ This similarity can be made explicit considering
the $n$-soliton form factors for the restricted sine-Gordon theory at the
reflectionless points in formula (31) of \cite{BabBS96}. Then, for the local
fields interpreted as primary operators in \cite{BabBS96}, the corresponding
form factors can be easily rewritten as determinants of $n\times n$ matrices
whose elements are integral convolutions of $n$-soliton wave functions (the $%
\psi $-functions (32)) plus contributions coming from the local operators.

The rough picture that seems to emerge is that by performing the IR limit on
our lattice form factors the lattice wave functions factorized in terms of
Q-operator eigenvalues have to converge to the infinite volume $n$-soliton
wave functions. To which extent this picture can be confirmed and clarified
by a detailed analysis of the thermodynamic limit starting from our lattice
sine-Gordon model results is of course an interesting question to which we
would like to answer in the future.

\subsection{Outlook}

It is worth mentioning that we didn't succeed yet to express the matrix
elements of discretized exponential of the sine-Gordon field in terms of one simple
determinant formula. Hence our next natural project is the simplification of
the present representation; this is also important in view of the attempt to
extend our results from the lattice to the continuous finite and infinite
volume limits. The main goal here is to derive the known form factors of the
IR limit, starting from our lattice form factors, in this way solving the
longstanding problem of the identifications of local fields in the S-matrix
characterization of the infinite volume sine-Gordon model.

Beyond the sine-Gordon model we want to point out the potential generality
of the method we have introduced here to compute matrix elements of local
operators for quantum integrable models. The main ingredients used to
develop it are the reconstruction of local operators in Sklyanin's SOV
representations and the scalar product formulae for the transfer matrix
eigenstates (and general separate states). The emerging picture is the
possibility to apply this method to a whole class of integrable quantum
model which were not tractable with other methods. This is in particular 
the case for lattice integrable quantum models to which the algebraic Bethe
ansatz does not apply. The first remarkable case is given by the $\tau_2$%
-model in general representations which are of special interest for their
connection to the chiral Potts model. This will be the next model that we
will analyze by our technique due to the similarity of its cyclic
representations with those of the sine-Gordon model.

There are also many other examples which are interesting and for which, on
the one hand, the reconstruction of the local operators can be deduced from 
\cite{MaiT00} and, on the other hand, the description of the spectrum can be
given by Sklyanin's quantum separation of variables. For all these models
the possibility to apply our method for the computation of matrix elements
is very concrete and moreover the results are expected to have a completely
similar form to the ones shown in the present article.

\bigskip

\textbf{Acknowledgments}

We would like to thank N. Kitanine, K. K. Kozlowski, B. M. McCoy, E.
Sklyanin, V. Terras and J. Teschner for their interest in this work. J. M.
M. is supported by CNRS. N. G. and J. M. M. are supported by ANR grant
ANR-10-BLAN-0120-04-DIADEMS. G. N. is supported by National Science
Foundation grants PHY-0969739. G. N. gratefully acknowledge the YITP
Institute of Stony Brook for the opportunity to develop his research
programs and the privilege to have stimulating discussions on the present
paper and on related subjects with B. M. McCoy. Moreover, G. N. would like
to thank the Theoretical Physics Group of the Laboratory of Physics at the
ENS Lyon and the Mathematical Physics Group at the IMB of the University of
Dijon for their hospitality.

\end{document}